\documentclass{article}
\usepackage[a4paper, total={6.5in,9in}]{geometry}
\usepackage[utf8]{inputenc}
\usepackage{amsmath}
\usepackage{amssymb}
\usepackage{graphicx}
\usepackage{authblk}
\usepackage{url}
\usepackage{natbib}
\usepackage{epstopdf}
\usepackage{appendix}
\usepackage{rotating}
\usepackage{comment}
\usepackage{appendix}
\usepackage[ruled]{algorithm2e}
\usepackage[english]{babel}
\usepackage[autostyle]{csquotes}

\vspace{-40pt}
\title{Topologically Mapping the Macroeconomy}
\vspace{-30pt}
\author[1]{Pawe{\l} D{\l}otko\thanks{Full Address: Mathematics Department, College of Science, Swansea University, Bay Campus, Swansea, SA1 8EN, United Kingdom. Email:p.t.dlotko@swansea.ac.uk. }}
\affil[1]{Mathematics Department, Swansea University, United Kingdom}
\vspace{-40pt}

\vspace{-30pt}

\author[2]{Simon Rudkin \thanks{\textbf{Corresponding Author}. Full Address: Economics Department, School of Management, Swansea University, Bay Campus, Swansea, SA1 8EN, United Kingdom. Tel: +44 (0)1792 606325 Email:s.t.rudkin@swansea.ac.uk}}
\affil[2]{Economics Department, Swansea University, United Kingdom}
\vspace{-30pt}
\author[3]{Wanling Qiu\thanks{Full Address: Accounting and Finance Subject Group, School of Management, University of Liverpool, 20 Chatham Street, Liverpool, L69 7ZH, United Kingdom. Email:wanling.qiu@liverpool.ac.uk}}
\affil[3]{School of Management, University of Liverpool, United Kingdom}
\vspace{-20pt}
\begin{document}
\maketitle
\begin{abstract}
An understanding of the economic landscape in a world of ever increasing data necessitates representations of data that can inform policy, deepen understanding and guide future research. Topological Data Analysis offers a set of tools which deliver on all three calls. Abstract two-dimensional snapshots of multi-dimensional space readily capture non-monotonic relationships, inform of similarity between points of interest in parameter space, mapping such to outcomes. Specific examples show how some, but not all, countries have returned to Great Depression levels, and reappraise the links between real private capital growth and the performance of the economy. Theoretical and empirical expositions alike remind on the dangers of assuming monotonic relationships and discounting combinations of factors as determinants of outcomes; both dangers Topological Data Analysis addresses. Policy-makers can look at outcomes and target areas of the input space where such are not satisfactory, academics may additionally find evidence to motivate theoretical development, and practitioners can gain a rapid and robust base for decision making. 
\end{abstract}

\maketitle
\section{Introduction}

Relationships within economics are often assumed monotonic, or to possess single peaked shapes with global optima, such that through econometric modelling may recover the correspondence. Use of regressions to extract causality from independent variable(s) to dependent is a fundamental in a statistics education. However, in the pursuit of parsimony many models are either products of a specific small sample, or fail to truly capture the relevant input data. Recognition of the need for simultaneous consideration of multiple dimensions has spawned a literature on high dimensional treatments \citep{belloni2014high} and the high dimensional econometrics R package \textit{hdm} \citep{chernozhukov2016hdm}. But, rather than this pursuit of regression techniques for short and very fat panels, it is in Machine Learning where much of the research effort currently lies. Where Machine Learning can help us construct data from novel sources in support of economic understanding \citep{mullainathan2017machine,athey2018impact} it still suffers from the pre-assumption of a search term. Be it the linear relationship of ordinary least squares, or the potentially over-fitted models from machine learning, empirical economics is still bound towards parsimony and following the lead of the researcher.

Topological data analysis (TDA) is a direct response to the challenge, with the intention to drive modelling upward from the data instead of downward from pre-defined search terms. Understanding should begin with a data cloud and the toolbox of TDA should then provide the means to formalise that appreciation for the data's contents. TDA captures the shape of the cloud, the topoogical features within it, and quantifies these for further analysis. From any point in a cloud, searches are undertaken to identify neighbours, grouping the data according to its location in multidimensional space. In this way TDA is like a clustering analysis, but unlike clustering approaches it preserves all of the topological features of the dataset.  \cite{carlsson2009topology} seminal work introduced the concept of persistent homology, an example of TDA in its' recognition of features that ``persist'' over large filtrations, provides the mathematical support to this intuition. Intrinsically TDA recognises data as random realisations from a surface and so, unlike approaches that view data as being drawn from specific distributions, it is robust to perturbations of the population. It has a robustness to noise, and continues to capture the inter-relationships between points no matter how the surface is manipulated. 

The strengths of TDA see it widely adopted in testing materials \citep{lee2017quantifying,buchet2018persistent}, gene mutations and medicine \citep{nicolau2011topology,xia2015multidimensional,patania2019topological}, and for detecting failure in systems \citep{guo2017identification,carlsson2018fibres}. All of these speak directly to Economics and this paper represents a first step in making that transition. There is also a growing literature on the topology of time series \citep{perea2015sliding,perea2019topological,pereira2015persistent,gidea2018topological}, which itself speaks to macroeconomic data. \cite{dlotko2019topologya} demonstrates how the periodicity detection of \cite{dlotko2019period} can be used to augment the yield spread in producing more accurate recession forecasts. This paper does not speak directly to that literature, but the demonstration of overlap with time series analysis through the examples presented show the contribution TDA Ball Mapper can make in summarising time series. As this literature continues to grow the need for economists to remain on top of the potential is strong.

From an economic perspective the potential of TDA, and particularly visualisation techniques like TDA Ball Mapper exposited here, lies in the way that it informs on those parts of the point cloud that other approaches miss. For example a regional economy may have the fiscal stimulii for growth, the education level amongst its residents and the openness for trade, but it may fail to capitalise upon its potential to anything like the extent that the three characteristics would suggest it should. A fourth, and seemingly unrelated by regression, characteristic may be its connectivity to the internet. Should the point in the datacloud be low in this dimension then the resulting low outcome could be explained, but the insignificance of the internet connectivity variable would have diverted policy makers attention long before any detailed look was taken at that region. In fact a quick win could be achieved there to provide the connectivity, and possibly to do so whilst reducing expenditure on the stimulus, such that the outcomes for the region improve. This is a simplistic example but motivates the reasoning why TDA Ball Mapper can become an integral part of summarising data.  

First and foremost this is a methodological paper which demonstrates, tests and adapts a contemporary data analysis approach from the mathematical sciences for use with economic data. The examples presented are illustrative rather than exhaustive reviews of their fields, but can be taken as launch points for further, and more data rich, explorations. Three key contributions to the literature are made. Firstly, the introduction of the newly formulated TDA Ball Mapper algorithm informs of a valuable technique for data summary. Secondly, measurement within the data cloud informs further on dynamics, including the distances the economy stands from its' worst realisations. Finally, the ability of TDA Ball Mapper visualisations to reduce complex systems into something which can be readily reduced and understood by practitioners facilitates a wealth of possibilities in forecasting. These three contributions are showcased through common macroeconomic examples as a first step to developing the appreciation of TDA for economic analysis.    

The remainder of the paper proceeds as follows. Section \ref{sec:tda} provides a theoretical overview of the TDA Ball Mapper approach \citep{dlotko2019ball} and how it nests within the wider discussion of TDA. An artificial panel data discussion follows in Section \ref{sec:art2}, with a review of the role of correlation. Two macroeconomic examples are exposited to showcase the benefits of using TDA. Firstly, Section \ref{sec:er} builds on \cite{jorda2019global} historic dataset to evaluate the evolution of leading economies. Secondly, Section \ref{sec:tdays} constructs TDA Ball Mapper graphs of the moments of private capital growth; highlighting complex relationships missed in the existing regression based literature. From the former emerges a discussion of the relationship between the contemporary economy and that of the Great Depression, while the latter offers a new take on the role of financial crashes in growth. Further opportunities for analysis, and a roadmap for the future analysis of the macroeconomy are given in Section \ref{sec:summary}.  

\section{TDA Ball Mapper}
\label{sec:tda}

A new field of data science, TDA, opens many avenues for gaining new understanding. The TDA Ball Mapper algorithm is developed by \cite{dlotko2019ball} from building blocks in Reeb and Mapper graphs\footnote{See \cite{carriere2018structure} for a further review of the original mapper software.}, and in persistent homology \citep{carlsson2009theory,carlsson2009topology}. Owing to the fundamental nature of these blocks a brief oversight is offered here with the exposition following the structure in \cite{dlotko2019ball}. 

\subsection{Persistent Homology and the Mapper Algorithm}

Consider the \textit{abstract simplical complex} $\mathcal{K}$, a set of sets such that for $s \in \mathcal{K}$ and $t \subset s$, $t \in \mathcal{K}$. For a point cloud $X$ a simple \textit{Vietoris-Rips} (VR) representation can be constructed from $n$ balls centered on $x_1,x_2,...,x_n \in X$ provided the balls around these points overlap; that is $B(x_i,r) \cap B(x_j,r)$ for balls $B()$ of radius $r$ centered on $x_1$ and $x_2$. This complex is then referred to as $VR(x,r)$. When $r$ increases so the complex with smaller radius must be included with that of larger radius. A filtration is born of the process of changing $r$. A sequence of inclusions in groups $H_p()$ can be written as $f_{r,r'} : H_p(VR(X,r)) \rightarrow H_p(VR(X,r'))$; persistent homology groups are the images of $f_{r,r'}$. Persistence intuitively can be thought of as existence of a set of features across a range of filtration radii.  

\cite{dlotko2019ball} distinguishes the original mapper functions from TDA Ball Mapper using the initials CM and BM respectively; this paper does likewise. Consider a real function $f: \mathcal{M} \mapsto \mathcal{R}$ on manifold $\mathcal{M}$. With the data cloud $X$ a pair of points $x,y \in \mathcal{M}$ have equivalence, $x ~ y$, if $f(x)=f(y)$ and $x$ and $y$ are in the same connected component of $f^{-1}(x)=f^{-1}(y)$. In this way CM, as introduced in \cite{singh2007topological}, draws inspiration from a \textit{Reeb graph}.

Formally consider a finite set $X$ sampled iid from a manifold $\mathbb{M}$. In such cases $f^{-1}(x)=\emptyset$; inverse images of overlapping intervals replace the inverse images of the points. These intervals $I_1,...,I_n$ then cover the range of $f$. The CM algorithm considers clusters in $f^{-1}(I_k) \subset X$ whereby the collection of clusters for each interval correspond to the vertices of the mapper graph. An edge is drawn between two points iff $C_1 \subset f^{-1}(I_k)$ and $C_2 \subset f^{-1}(I_{k+1})$ satisfy $k \in \lbrace 1,...,n-1\rbrace$ and $C_1 \cap C_2 \neq \emptyset$. In such cases $C_1$ and $C_2$ are connected components, or clusters. Each cluster is represented by the point at its core. Edges are added for all pairs of points where there is data in the intersection of the corresponding clusters. 

The CM algorithm has four key inputs. Firstly the user must specify the data, ensuring any transformation is applied before supplying the input to the TDA code. Secondly the function that maps the data onto the outcome, $f : X \mapsto \mathbb{R}$ should be specified. Thirdly the cover of $\mathbb{R}$ should be specified, including the percentage of overlap needed to warrant a connection. Finally the user specify a clustering algorithm. Effective use of CM requires experience of what works in the algorithm as well as knowledge of the data being studied. TDA Ball Mapper as detailed in \cite{dlotko2019ball} requires just one parameter. Output from CM remains \textit{an overlapping cover} $C$ of the point cloud $X$.

\subsection{TDA Ball Mapper Algorithm}

TDA Ball Mapper (BM) generates a cover $C$ of the point cloud $X$ by generating a set of balls $B(C) \bigcup_{x \in C} B(x,\epsilon)$ which cover the entire set X. There are many ways to select the points that will form the centre of the balls, here an $\epsilon$-net is used. $\epsilon$ becomes the parameter of the algorithm. Algorithm 1 of the \cite{dlotko2019ball} paper identifies neatly how the net may be formed.

\begin{algorithm}[H]
	\SetAlgoLined
	\KwIn{Point cloud $X$, filtration parameter $\epsilon>0$}
	Identify uncovered points in $X$\;
	Create an intially empty cover vector $B\left(X,\epsilon\right)$\;
	\While{Any element of X uncovered}{
		Pick uncovered point, $p \in X$\;
		For every point in $x \in B(p,\epsilon)$ add p to cover $B(X,\epsilon)[x]$\;
		}
	\KwOut{$B(X,\epsilon)$}
	\caption{Greedy $\epsilon$-net\citep{dlotko2019ball}}
\end{algorithm}
\begin{algorithm}
	\SetAlgoLined
	\KwIn{$B(X,\epsilon)$}
	V = cover elements in $B(X,\epsilon)$ and $E = \emptyset$, \;
	\For{$p \in X$}{
		For each pair $c_1,c_2$ in $B(X,\epsilon)[p]$ add edge between vertices corresponding to the cover elements $c_1,c_2$.\;
		Formally $E =E\cup\lbrace c_1,c_2\rbrace $ 
	}
	\KwResult{BM Graph, G=(V,E)}
	\caption{Construction of a BM graph \citep{dlotko2019ball}}
\end{algorithm}
    
Conversion of the output from Algorithm 1 into a TDA Ball Mapper graph requires Algorithm 2\footnote{This is Algorithm 3 in \cite{dlotko2019ball}.}. Loops may be readily added to cover multiple filtration parameters, and this is implemented in the  \textit{BallMapper}\citep{dlotkor} R code. BM is expected to provide a higher number of vertices than CM, with the result that there may be additional information visualised in the graph. However, the TDA mapper does not represent with any sense of scale meaning that points which are actually close, but not connected. 

A primary use for TDA Ball Mapper is as a summary visualisation to guide further investigation. In this way the colouration function applied on the balls is important. By observing variation in outcome across the cover interesting cases may identify themselves. For example in an otherwise high outcome area of the characteristic space there may be low outcomes linked to just one of the dimensions. In such cases most monotonic functions would suggest that group should have a high outcome because of the characteristics which do align. This is seen in context in the applications. The user is cautioned to consult summary statistics for the balls to inform their conclusions.

\subsection{TDA Ball Mapper in R \citep{dlotkor}}

Newly released code for R \citep{cranr}, \textit{BallMapper} \citep{dlotkor} offers a series of useful tools for undertaking TDA Ball Mapper analyses. These tools are exposited more fully in the accompanying package documentation\footnote{At the time of writing this draft paper the package documentation was still in development. This will be released in mid-October.}. This paper makes use of two additional functions. Firstly the ability to colour the BM graph using the values of the different input axes instead of the outcome of the function. In this way it is possible to understand where in the abstract plot corresponds to the highest (lowest) values of each axis. Secondly the \textit{BallMapper} coding enables the user to specify a reference set and has the BM graph coloured according to the overall distance between any given ball and the reference point. As used the distance is simply the sum of the absolute euclidean distances but can be implemented using any appropriate distance function. 

\subsection{Interpretation of TDA Ball Mapper Graphs}

\begin{figure}
	\begin{center}
		\caption{TDA Ball Mapper Graph Interpretation}
		\label{fig:coloreg}
		\begin{tabular}{c c c}
			\multicolumn{3}{c}{\includegraphics[width=10cm]{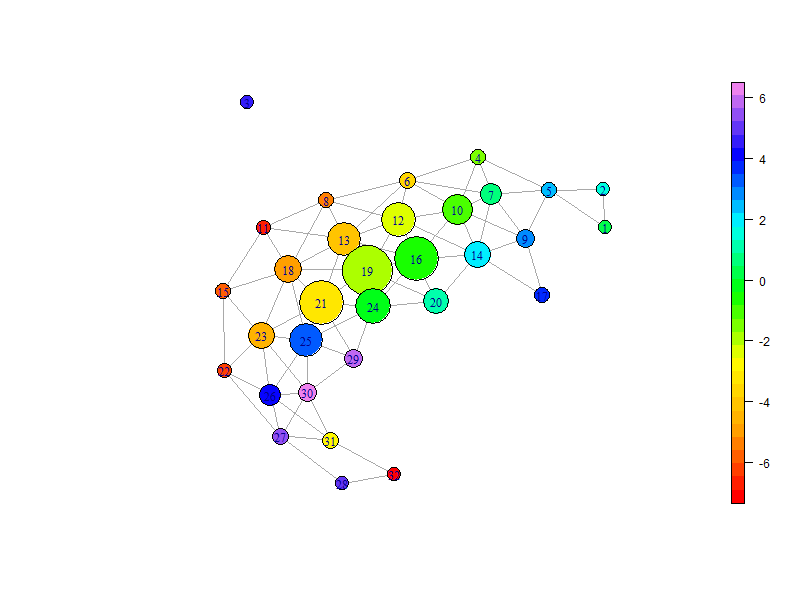}}\\
			\multicolumn{3}{c}{(a) TDA Ball Mapper plot}\\
			\includegraphics[width=4cm]{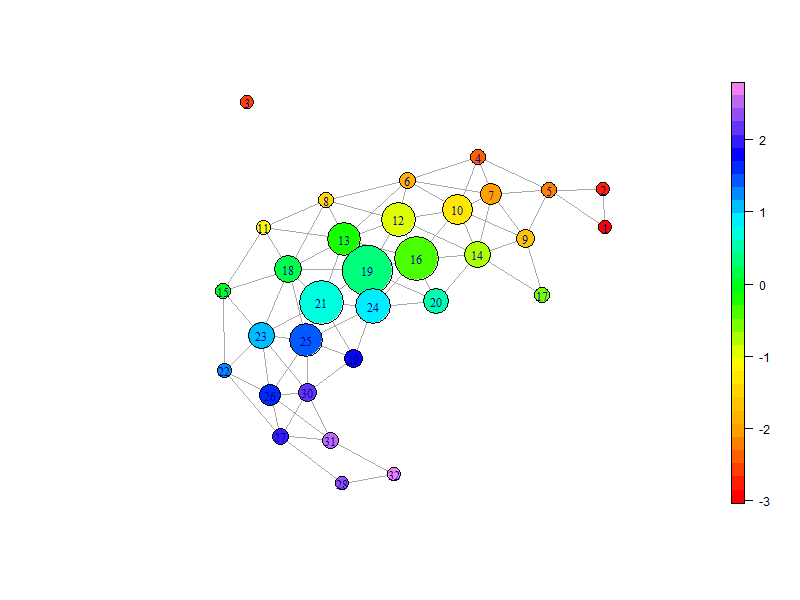}&
			\includegraphics[width=4cm]{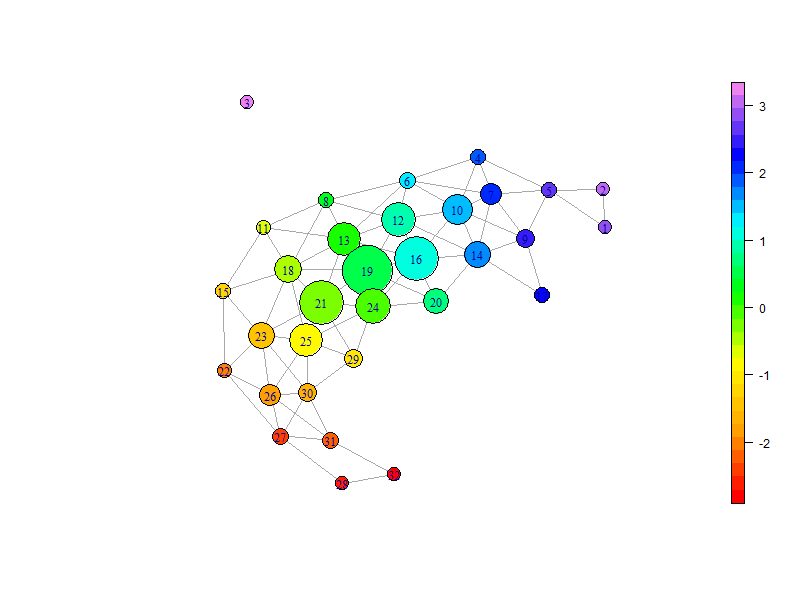}&
			\includegraphics[width=4cm]{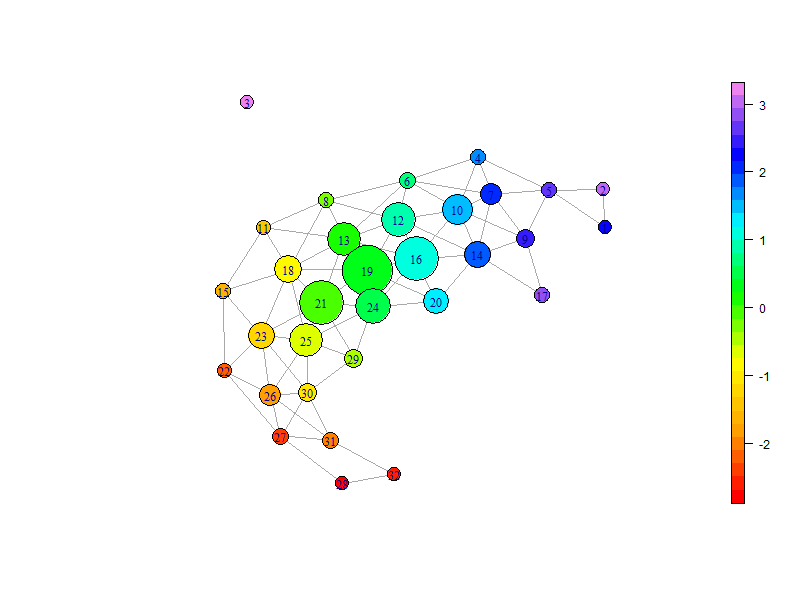}\\
			(b) Variable 1 & (c) Variable 2 & (d) Variable 3\\
		\end{tabular}
	\end{center}
\raggedright
\footnotesize{Notes: Artificial data TDA Ball Mapper example plotted using \textit{BallMapper} \cite{dlotkor}. Panel (a) coloured according to the outcome variable of interest. Panels (b) to (d) coloured by the respective characteristic variables.}
\end{figure}

Graphs produced from the \textit{BallMapper} package \citep{dlotkor} are analogous to those discussed in \cite{dlotko2019ball}. They are abstract representations of the underlying point cloud and may be coloured according to an outcome of interest. To illustrate how to interpret a TDA Ball Mapper graph consider the example in Figure \ref{fig:coloreg}. This plot is generated using three variables, two of which have strong negative correlations with the third. All are computed as random draws from a normal distribution, but are constructed to have fixed correlations. The outcome variable is only loosely correlated with the axes\footnote{The correlation matrix in this case is $\begin{bmatrix}
	1 & -0.830 & -0.660 & 0.165 \\
	-0.830 & 1 & 0.967 & 0.022\\
	-0.660 & 0.967 & 1 & 0.106\\
	0.165 & 0.022 & 0.106 & 1 \\
	\end{bmatrix}$. Rows 1 to 3, and columns 1 to 3, correspond to the axes. Row (column) 4 is the outcome variable.}. Correlation's role in the TDA Ball Mapper plots is returned to later. 

Figure \ref{fig:coloreg} shows a typical output from \cite{dlotkor}'s \textit{BallMapper} package. The main panel is the TDA Ball Mapper plot for the three artificial variables constructed for the illustration. In the picture the edges that join balls are clear, as is the colouration of the balls. A single ball is located to the upper left of the panel; such observations are outliers from the main dataset. When there are no edges connecting balls then it can be concluded that the points within those balls are sufficiently different from each other to warrant consideration as separate entities. In this way balls 1, 32 and 17 are all to be thought of as being in different parts of the point cloud. In TDA Ball Mapper diagrams the size of the ball gives an indication of the number of points within that ball. Many larger balls appear near the centre of the plot; an inevitable consequence of using randomly generated normally distributed variables in which the majority of the values will be close to the mean. 

Panel (a) does say little about what the values of the respective variables actually correspond with which part of the abstract representation. Hence it is helpful to colour the plot according to the axes of the point cloud. Panel (b) of Figure \ref{fig:coloreg} informs that the lowest values of the first variable are to be found at the top right of the plot, whilst the highest values can be found in the lower reaches of the plot. Broadly these two extremes are associated with high values of the outcome. A quadratic relationship between variable 1 and the outcome is implied. Variables 2 and 3 are highly correlated with each other, and have a strong negative association with variable 1, as can be seen in panels (c) and (d). Here again the high outcome values at both ends of the plot suggest that it is better to be in the extremes of the distributions of these two variables also. An initial association like this is a first step to interpreting the TDA Ball Mapper graph.

Secondly look for cases where the colouration does not follow the suggested relationship. For example being at the ends of the graph, the extremes of the distribution of the variables, is broadly linked to the best outcomes. However a closer look at panel (a) shows balls 31 and 32 to be interesting cases. Here the average colouration of the balls is amongst the lowest values despite their location in areas of the point cloud otherwise associated with high outcomes. In many applications it is the ability of the TDA Ball Mapper graphs to highlight cases like this that are the main attraction. Using the smaller plots, or by matching back to the main dataset and summarising the contents, it can be identified specifically which dimension is making the difference between success and failure at achieving high outcomes.  

\begin{figure}
	\begin{center}
		\caption{TDA Ball Mapper Graph Interpretation}
		\label{fig:coloreg2}
		\begin{tabular}{c c c c}
			\includegraphics[width=3.6cm]{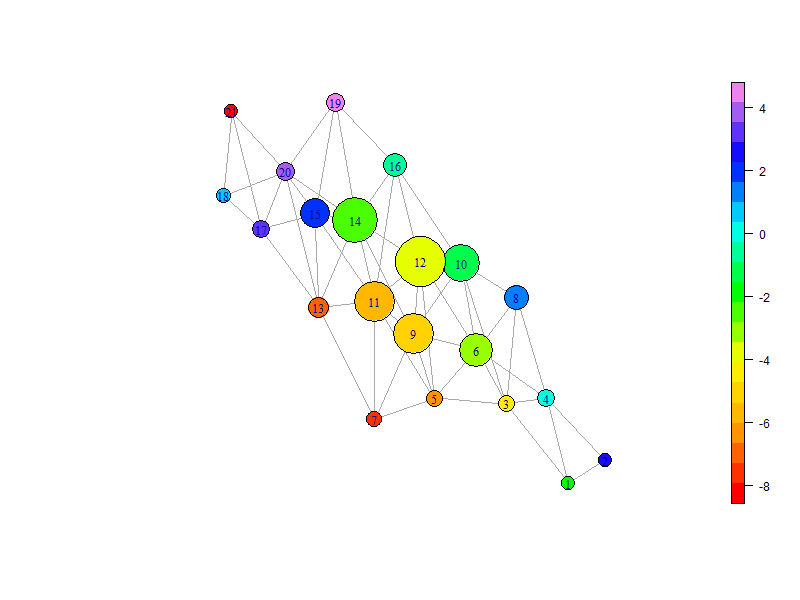} &
			\includegraphics[width=3.6cm]{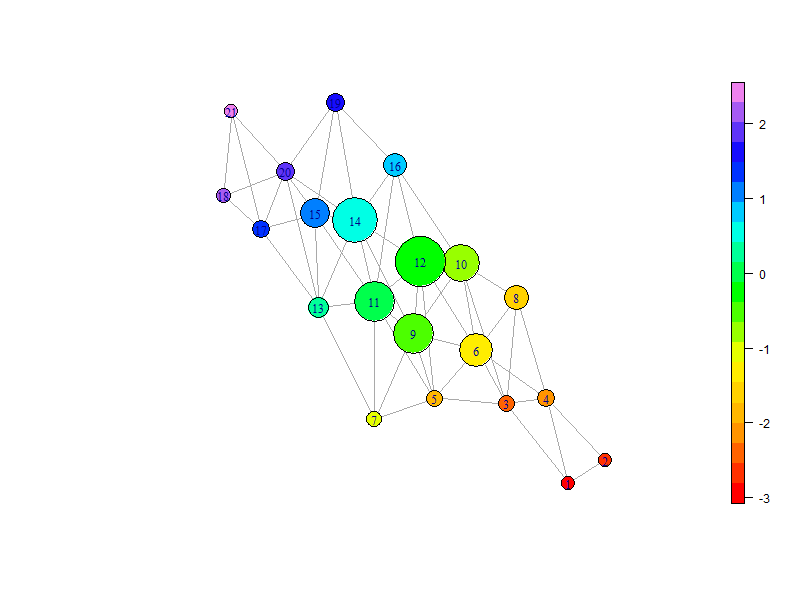}&
			\includegraphics[width=3.6cm]{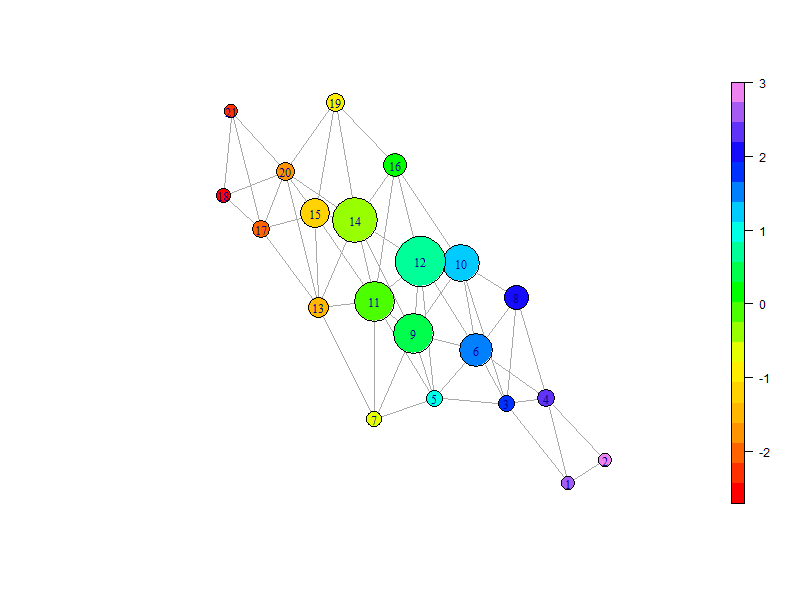}&
			\includegraphics[width=3.6cm]{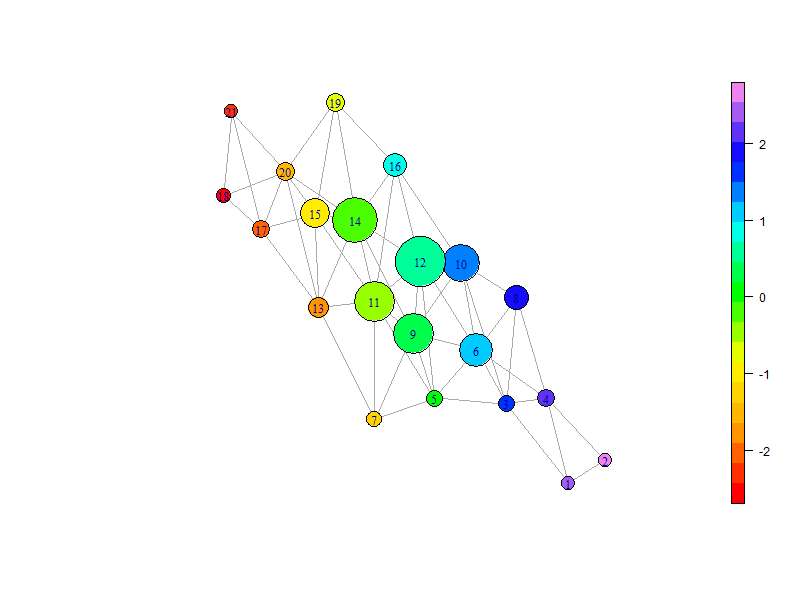}\\
			(a) TDA Ball Mapper plot & (b) Variable 1 & (c) Variable 2 & (d) Variable 3\\
		\end{tabular}
	\end{center}
	\raggedright
	\footnotesize{Notes: Artificial data TDA Ball Mapper example plotted using \textit{BallMapper} \cite{dlotkor}. Panel (a) coloured according to the outcome variable of interest. Panels (b) to (d) coloured by the respective characteristic variables.}
\end{figure}

TDA Ball Mapper has just one parameter, the level of filtration used to construct the graph. Computationally it is easy to use any level of filtration. Too large an $\epsilon$ will see all points become combined into a single ball; too low will see too few connections to make any meaningful inference. Another artificial example informs the discussion of filtration size. Figure \ref{fig:coloreg2} has the same three input variables, and the same colouration function, but now the filtration level is increased to 1.2. The number of ball is notably reduced, and the shape is simplified. Ball 3, intially an outlier in Figure \ref{fig:coloreg} becomes part of one of the connected balls leaving no outliers in Figure \ref{fig:coloreg2}. However, the high outcomes at the two ends of the shape are once again clear; as is the negative correlation between variable 1 and 2 (3). In the previous discussion an interesting case in balls 31 and 32 was identified. These balls can now be found in balls 1 and 3, but the difference in colouration is not as stark as it was in Figure \ref{fig:coloreg}. This illustrates neatly how the story around balls 31 and 32 can be lost amongst choosing a too high $\epsilon$.

Interpretation is thus summarised as a first understanding of the shape, a second exploration of the contribution of the axes of the point cloud, and a third analysis of any seeming anomalies within the outcome. Only one parameter is required, the filtration level $\epsilon$. Choice of $\epsilon$ is a trade off between reducing the number of balls to something manageable, seeking connectivity amongst the data, and ensuring that there are not important features getting subsumed within a higher level of aggregation. In this paper all examples are considered against this triple objective carefully with robustness carried out for other filtrations as appropriate. 

\subsection{Methodological Summary}

TDA Ball Mapper is a means for the production of abstract two-dimensional representations of multi-dimensional data. The algorithms of TDA Ball Mapper generate a cover for the data cloud, the cover then being illustrated with a TDA Ball Mapper Graph. Connectivity of balls is a confirmation that there is proximity in the characteristics space, but distances between points in the graph convey no meaning. Special cases can be quickly identified and investigated further, whilst functionality in the new R package \textit{BallMapper} \citep{dlotkor} facilitates a deeper exploration of what is shown. As presented the approach offers a more effective way to summarise data, identify cases of research interest and to target policy accordingly. Limits on what may be done with the approach stem primarily from the openness of data sources, and the willingness of the researcher to engage with the data cloud. The remainder of the paper builds the case for TDA, demonstrating its benefits in a artificial and contemporary datasets. Results from an extensive macroeconomic returns history database illustrate TDA Ball Mapper in action and offer a foretaste for what could be done in Economics.

\section{Artificial Examples}
\label{sec:art2}


In macroeconomics panel data is most common. \cite{breitung2015analysis} reviews the contrast between these macro panels and the microeconomic datasets more commonly studied with cluster based approaches. Big data, or more specifically the need to capture the digital economy \citep{coyle2019cloud}, embed an appreciation of measuring text \citep{thorsrud2018words}, and make use of all available data sources without over fitting drive towards more variables than time periods for each unit. Nowcasting models are defined by large panels, with the new big data sources ever expanding the set \citep{giannone2008nowcasting}. Notwithstanding the ongoing question of data reduction \citep{de2008forecasting}, or the dangers of integration in macroeconomic series \cite{banerjee2004some}, there is value in exploring how TDA Ball Mapper illustrates panel data \citep{hendry2018deciding}.

\subsection{Dataset Construction}

For the illustrations of this section it is necessary to control correlation and so a set of variables of known correlation is constructed. For a 1000 observation $x_0$ a matrix of random variables $\mathbf{A}$ is populated such that each column of the 1000 row matrix has known correlation with $x_0$. Production of the dataset begins with two variables, $x_0$ and $y_0$ which are both drawn randomly from the same distribution. The additional variable $x_i$, $i \in [1:198]$ is constructed such that:
\begin{align}
x_i = scale(x_0) * r  +  scale(residuals(lm(y_0~x_0))) * sqrt(1-r*r) \label{eq:cor}
\end{align}    
in which $r$ represents the desired correlation. To capture the full range $r$ is set as $\left(i/100 \right)-1$, such that the matrix $\mathbf{A}$ has columns with increments of correlation of 0.01. 

For output there are three series considered. In two cases the balls are coloured according to the average observation number found within. Because the data is unsorted colouring directly by observation number will produce something close, but not completely, random. By contrast sorting the data prior to generating the row id and performing colouration. A simple linear combination is applied, this would correspond to the case where there is a known theoretical relationship underpinning the observed values. Specifically the outcome variable $M_i$ is generated for observation $i$ as $M_i=0.3x_{0i}+0.6y_{0i}+\omega_i$, where $\omega_i$ is a normally distributed error term drawn independently from $N(0,1)$. As with previous time series a noise is added; this captures the difference between fitted values and those which would be seen in a real sample. 

\subsection{Role of Correlation}

\begin{figure}
	\begin{center}
	 	\caption{Correlation and Mapper}
	 	\label{fig:cor1}
	 	\begin{tabular}{c c c c}
	 		\multicolumn{4}{l}{$\rho=-0.9$:}\\
	 		\includegraphics[width=4cm]{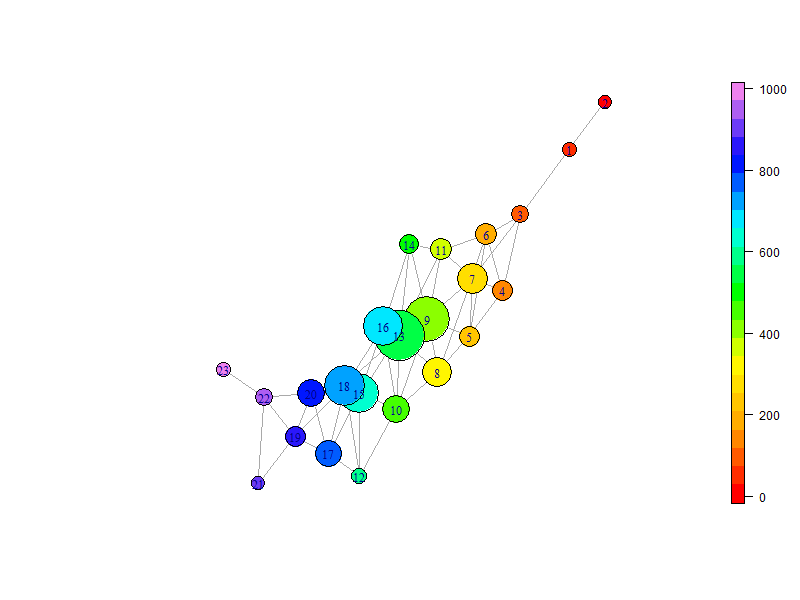}&\includegraphics[width=4cm]{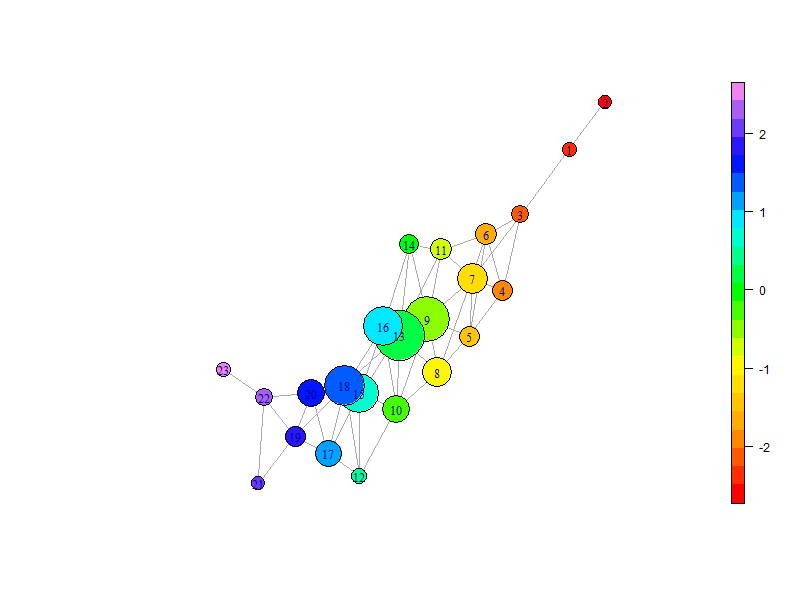}&
	 		\includegraphics[width=4cm]{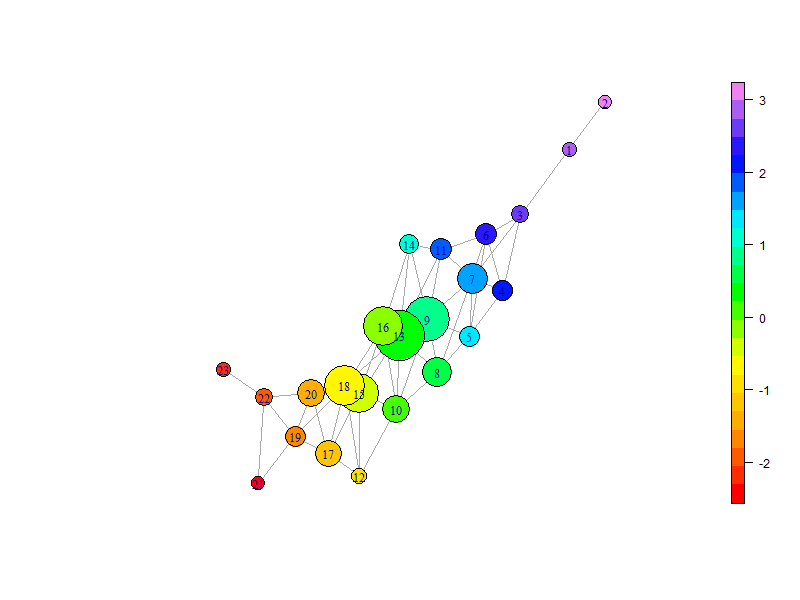}&\includegraphics[width=3cm]{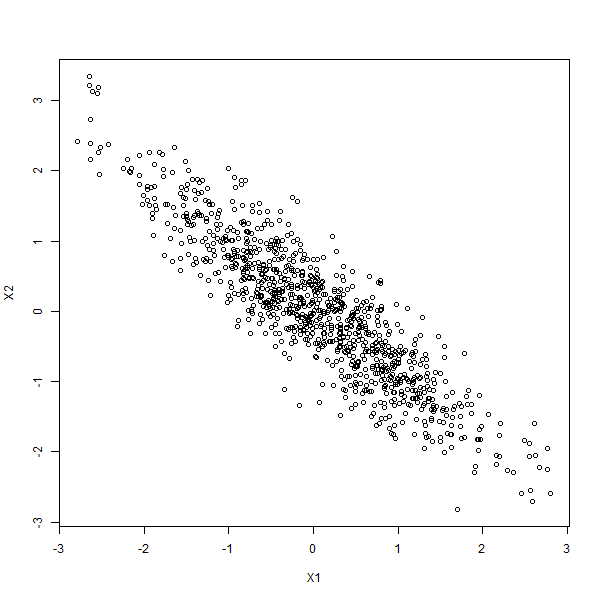}\\
	 		\multicolumn{4}{l}{$\rho=-0.6$:}\\
	 		\includegraphics[width=4cm]{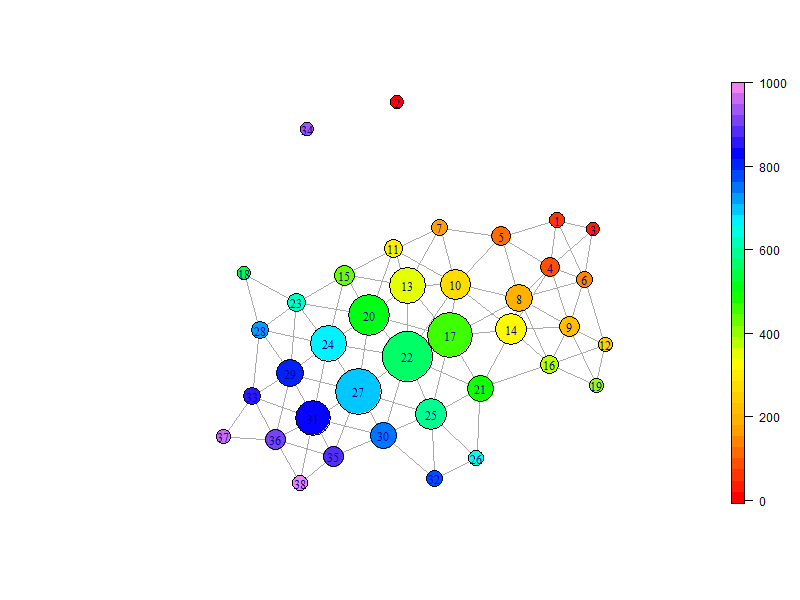}&\includegraphics[width=4cm]{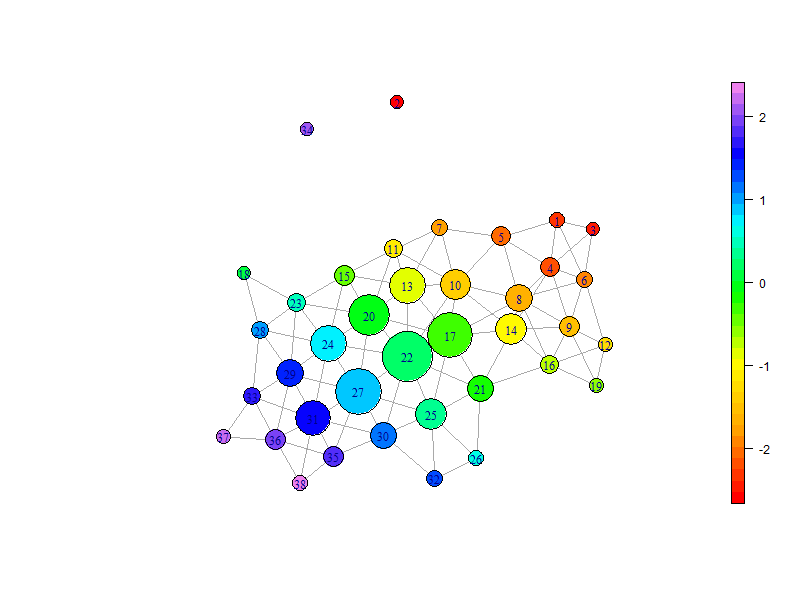}&
	 		\includegraphics[width=4cm]{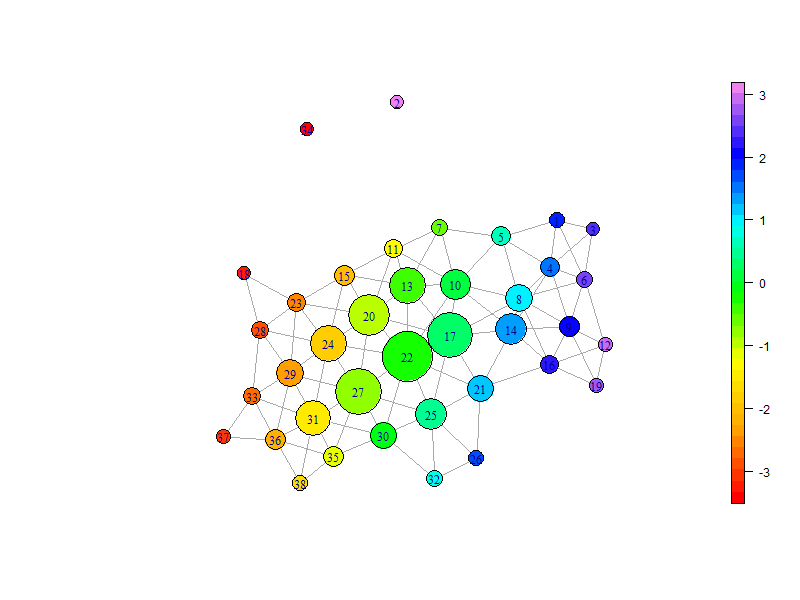}&\includegraphics[width=3cm]{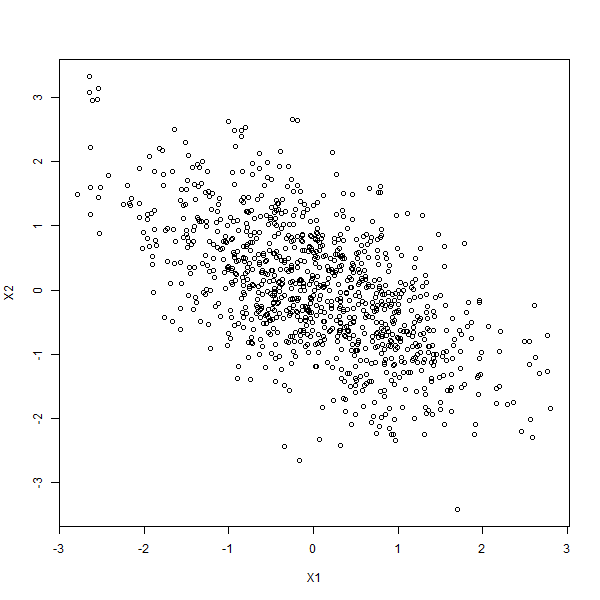}\\
	 		\multicolumn{4}{l}{$\rho=-0.3$:}\\
	 		\includegraphics[width=4cm]{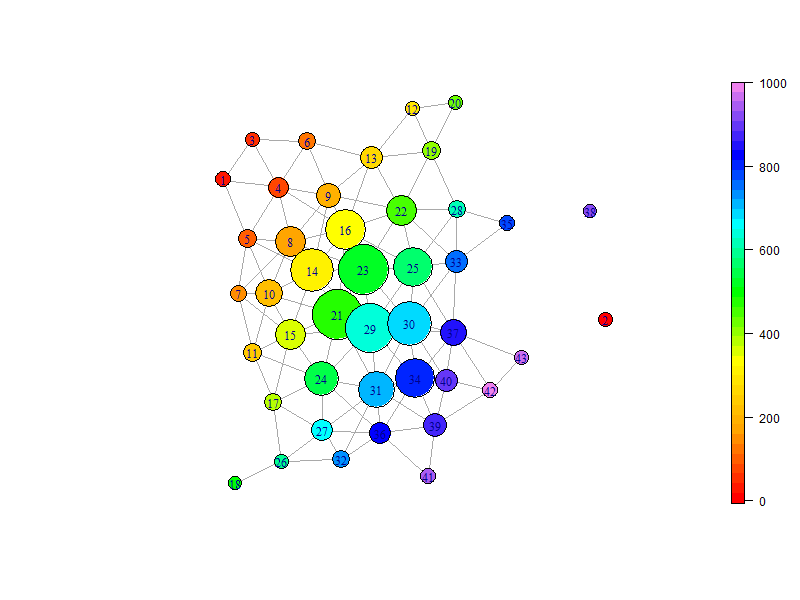}&\includegraphics[width=4cm]{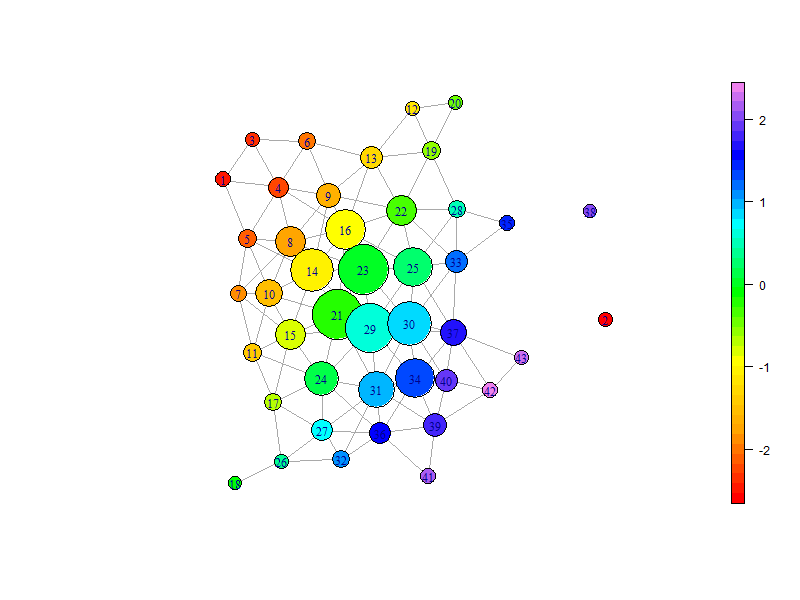}&
	 		\includegraphics[width=4cm]{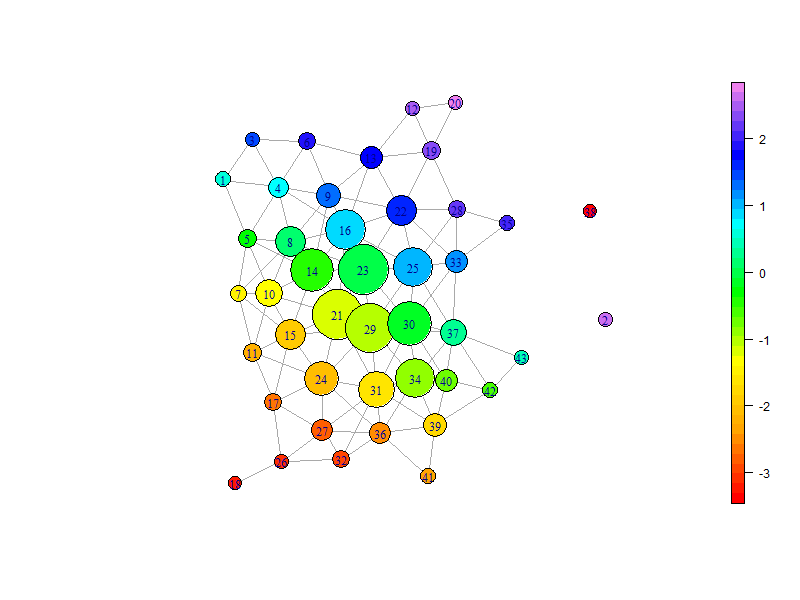}&\includegraphics[width=3cm]{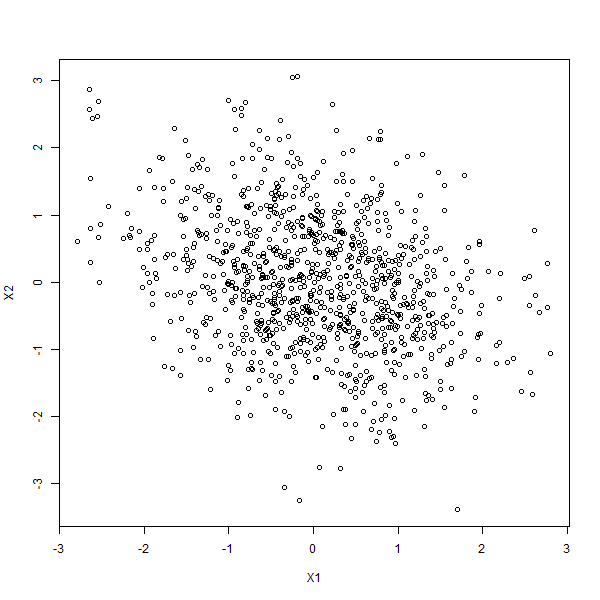}\\
	 		\multicolumn{4}{l}{$\rho=0.0$:}\\
	 		\includegraphics[width=4cm]{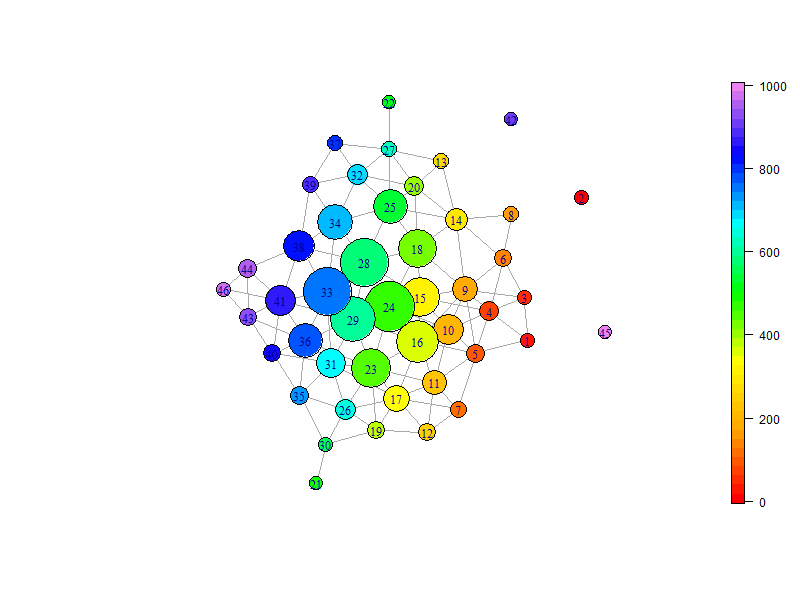}&\includegraphics[width=4cm]{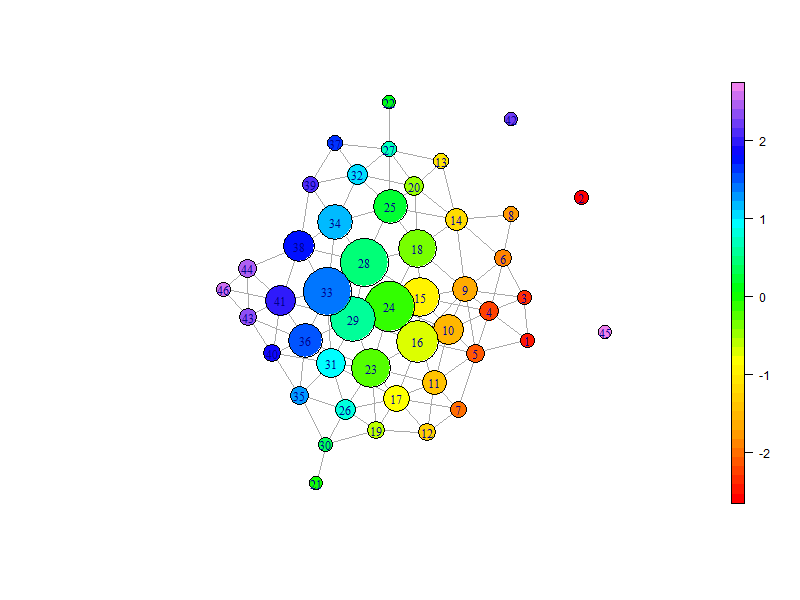}&
	 		\includegraphics[width=4cm]{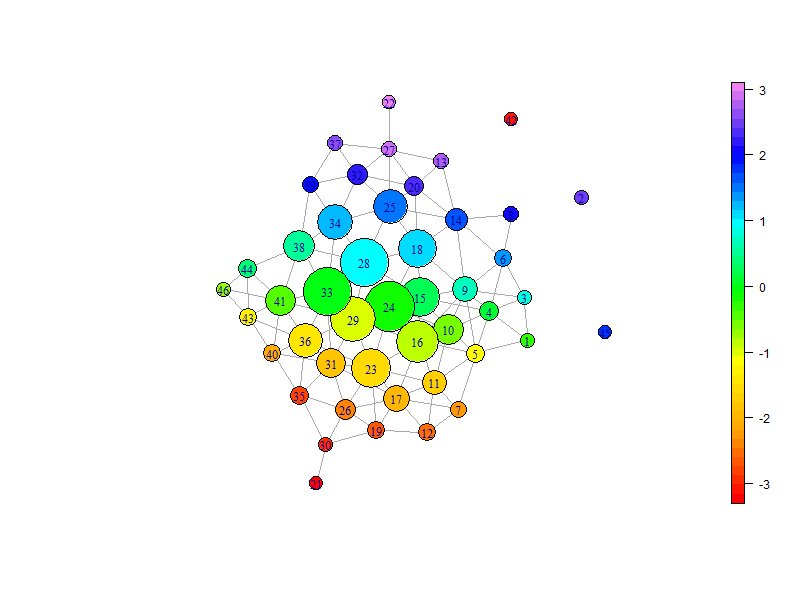}&\includegraphics[width=3cm]{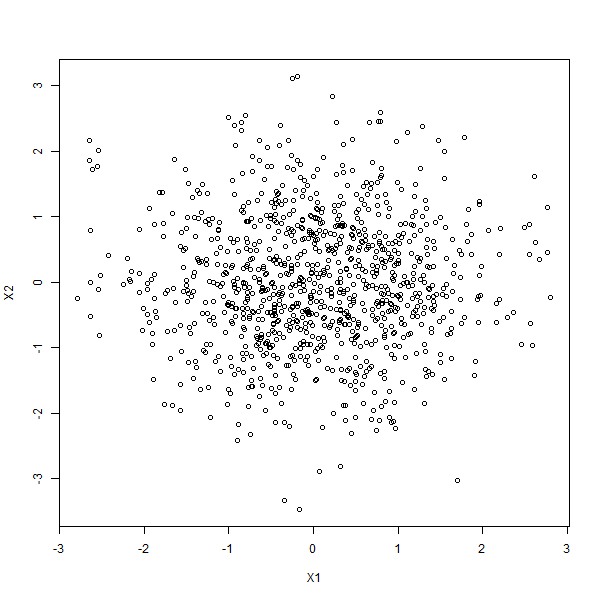}\\
	 		\multicolumn{4}{l}{$\rho=0.6$:}\\
	 		\includegraphics[width=4cm]{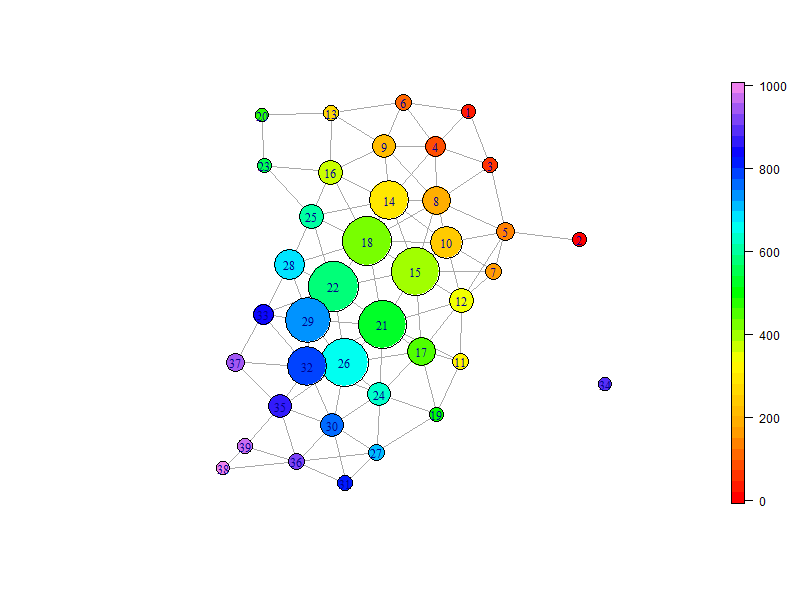}&\includegraphics[width=4cm]{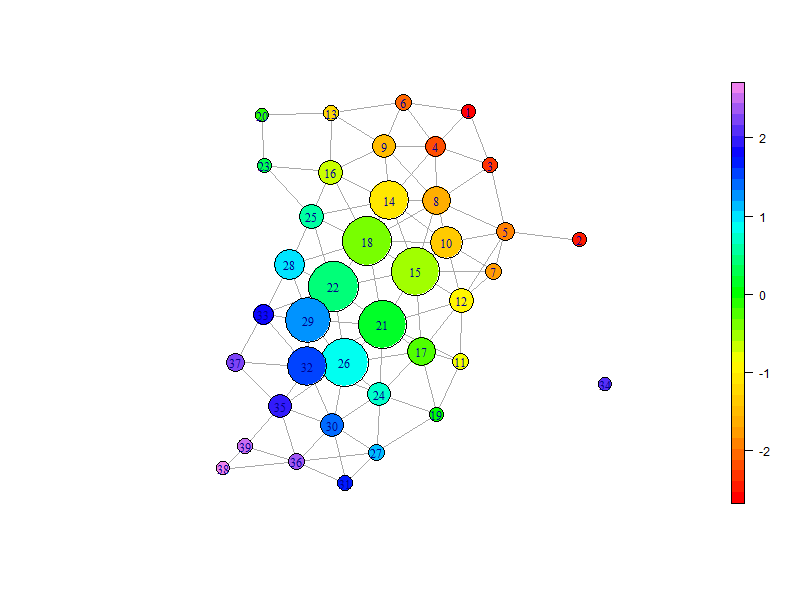}&
	 		\includegraphics[width=4cm]{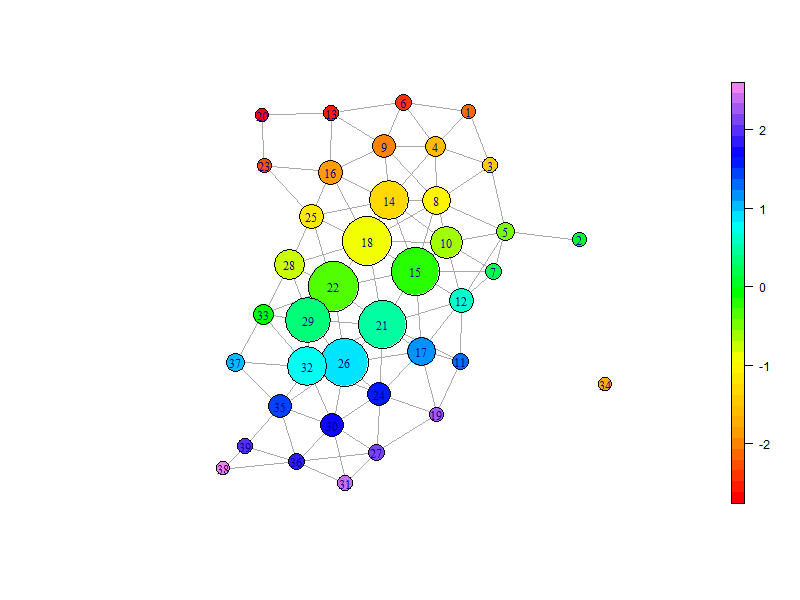}&\includegraphics[width=3cm]{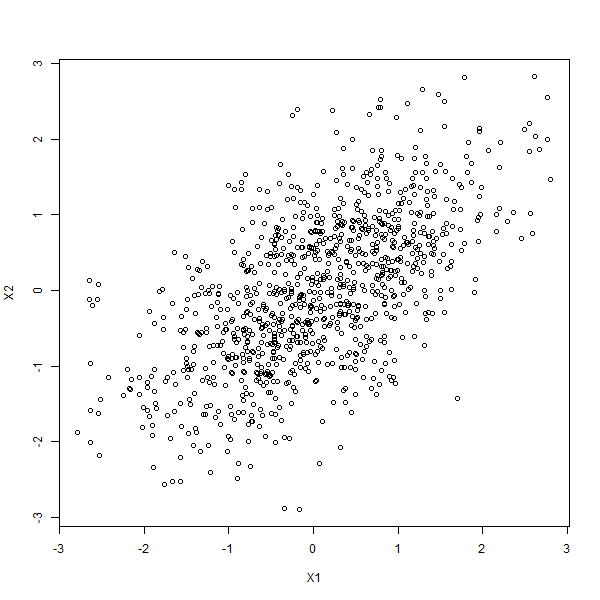}\\
	 		\multicolumn{4}{l}{$\rho=0.9$:}\\
	 		\includegraphics[width=4cm]{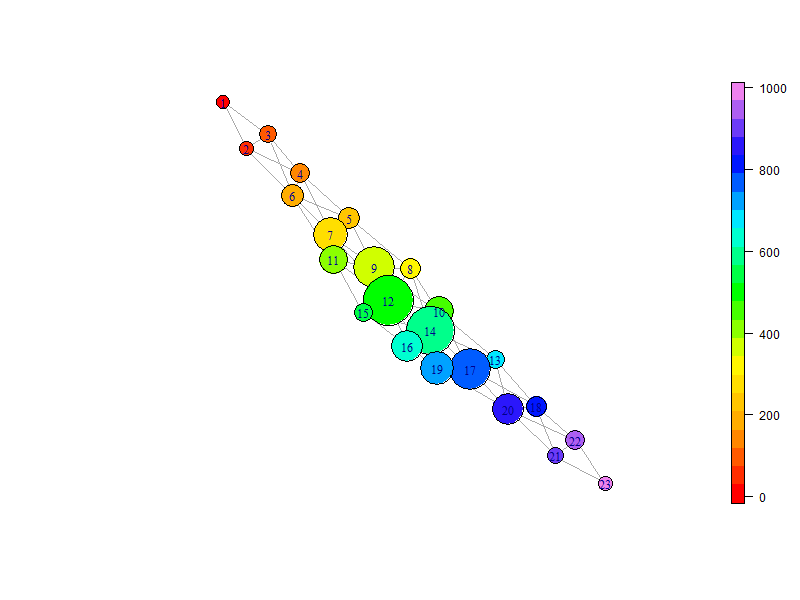}&\includegraphics[width=4cm]{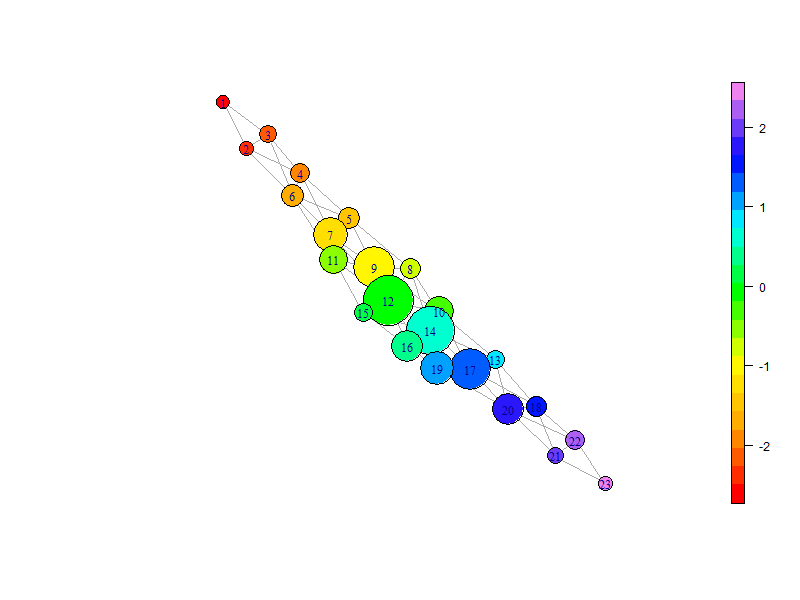}&
	 		\includegraphics[width=4cm]{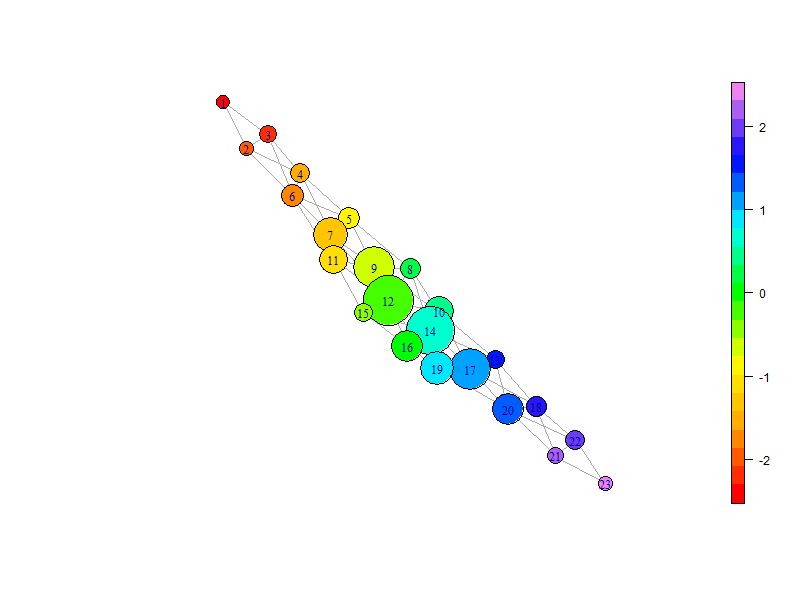}&\includegraphics[width=3cm]{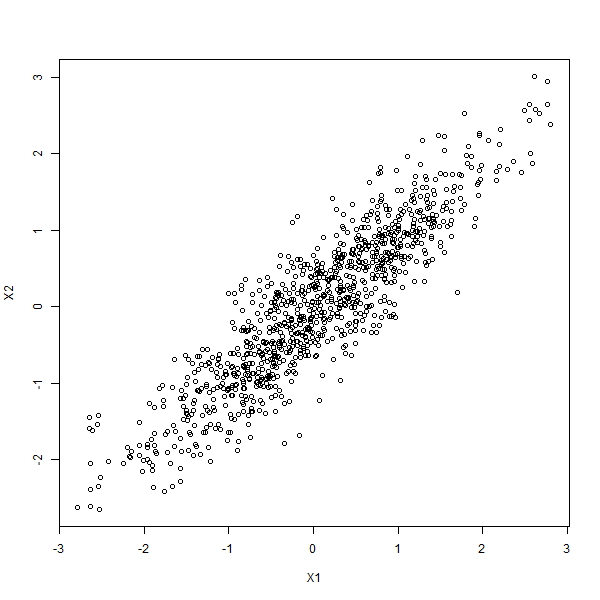}\\
	 	\end{tabular}
	\end{center}
\raggedright
\footnotesize{Notes: Left column shows mapper plot for a pair of bivariate normally distributed variables, $X$ and $Y$ with given correlation $\rho$. Observations are sorted on $X$ and coloured by the resulting observation number. In the centre two columns the graphs are coloured according to the two variables. In the right column a scatterplot of the data is provided. All TDA Ball Mapper plots generated using \textit{BallMapper} \citep{dlotkor} with filtration $\epsilon=0.7$.} 
\end{figure}

As a first exercise in understanding the way that correlation affects the shape of the TDA Ball Mapper plot consider Figure \ref{fig:cor1}. In total there are 5 different correlations shown, with each coloured in the left column according to the observation number, and in the centre two columns according to the input variables. The right hand column presents a scatter plot of the two variables. These are bivariate plots and as such it becomes immediately obvious that the TDA Ball Mapper algorithm is generating an abstract representation of the data. 

Working down from the top of Figure \ref{fig:cor1} the correlations are $\rho=-0.9$, $\rho=-0.6$, $\rho=-0.3$, $\rho=0.0$, $\rho=0.6$ and $\rho=0.9$. Because the data is sorted according to one of the variables there is a natural progression in the colouring from the lowest reds to the highest purples. Some similarities between the point clouds in the scatterplot and the TDA Ball Mapper diagrams is noted, but this is only seen in the two-dimensional plots as the next example demonstrates. These plots highlight a useful function in \textit{BallMapper} \citep{dlotkor}, whereby the graphs may be coloured according to their various axes. In the top row of Figure \ref{fig:cor1} the strong negative correlation means that in the bottom left of the diagrams, where $X$ is low, $Y$ is high. Columns 2 and 3 show this well. When the correlation falls, through the middle rows, the colouration no longer starts at opposite ends of the plot. Instead there is some commonality where both series have similar values in the same area of the TDA Ball Mapper graph. In the strong positive correlation, $\rho=0.9$ row at the bottom of Figure \ref{fig:cor1} the highest values of both $X$ and $Y$ are found in the same part of the plot.

\begin{figure}
	\begin{center}
		\caption{Trivariate Impact of Correlation}
		\label{fig:cor2}
		\begin{tabular}{c c c c}
			\includegraphics[width=3.6cm]{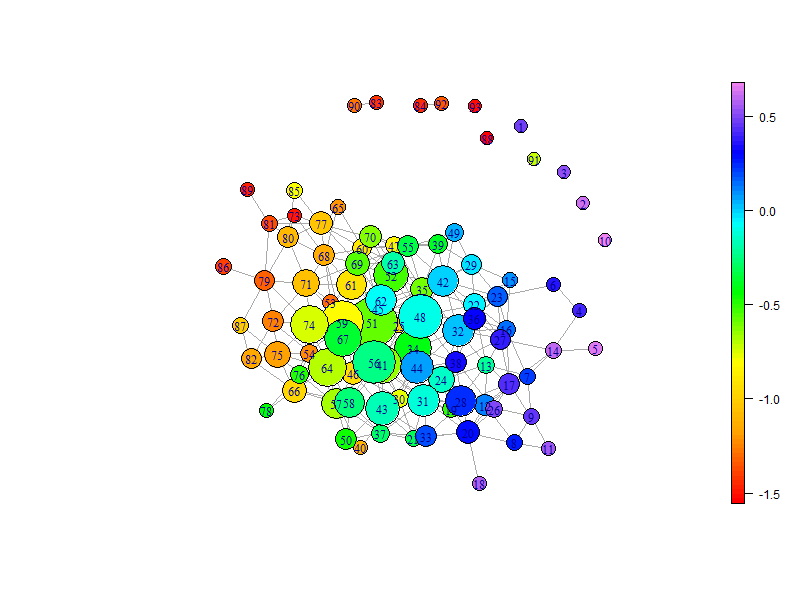}&\includegraphics[width=3.6cm]{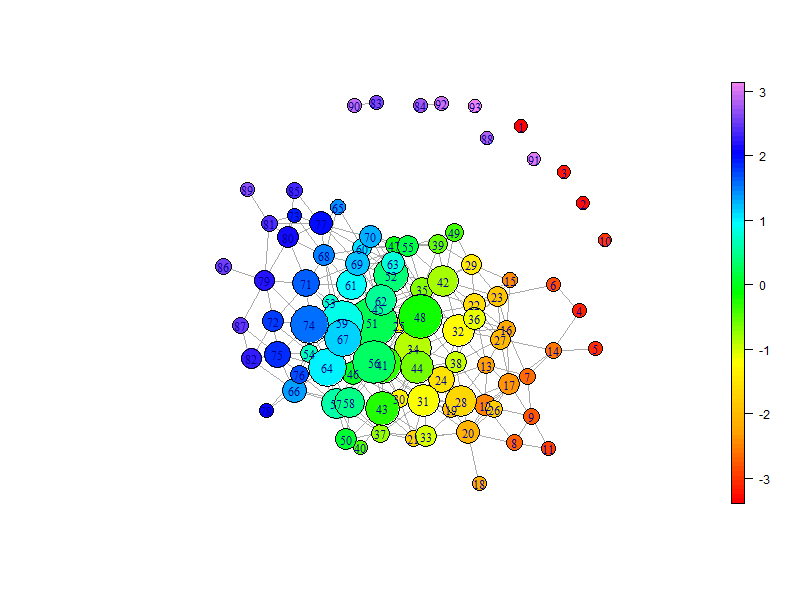}
			&\includegraphics[width=3.6cm]{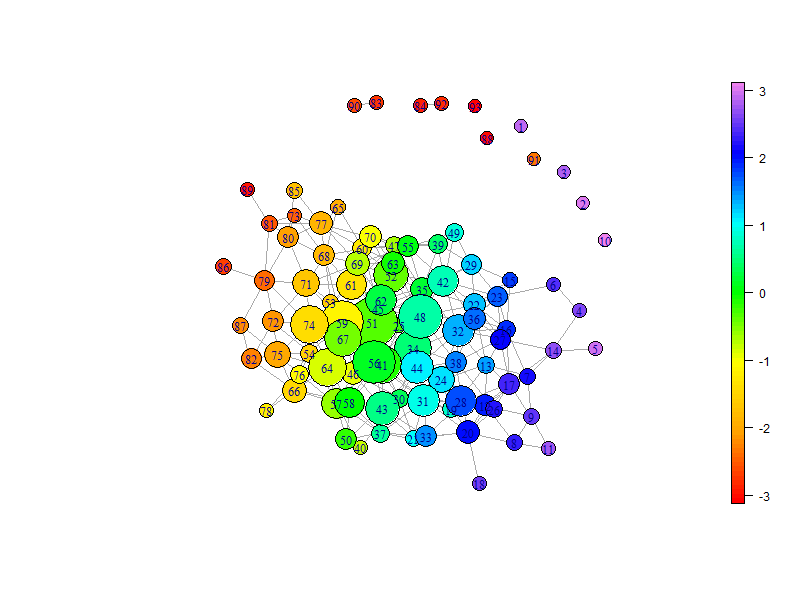}&\includegraphics[width=3.6cm]{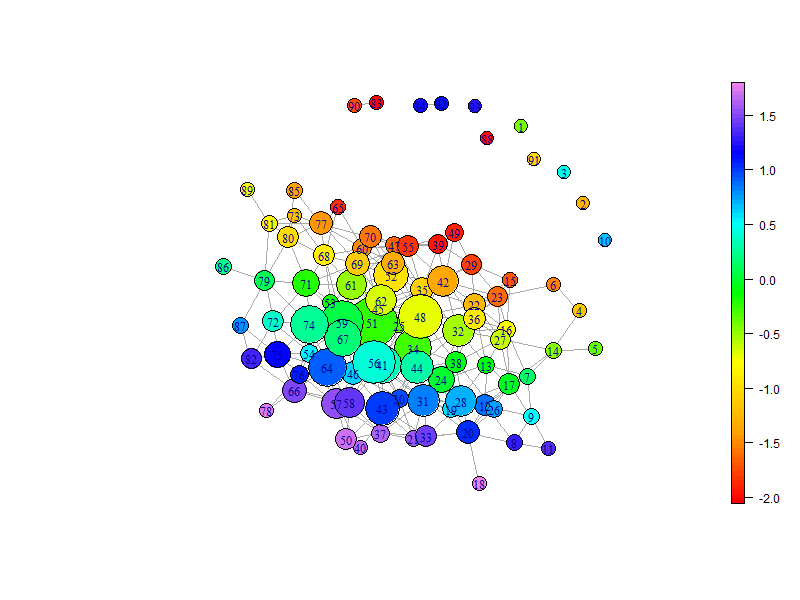}\\
			(a) Outcomes & (b) $x_0$ & (c) $y_0$ & (d) $z_0$ \\
			\includegraphics[width=3.6cm]{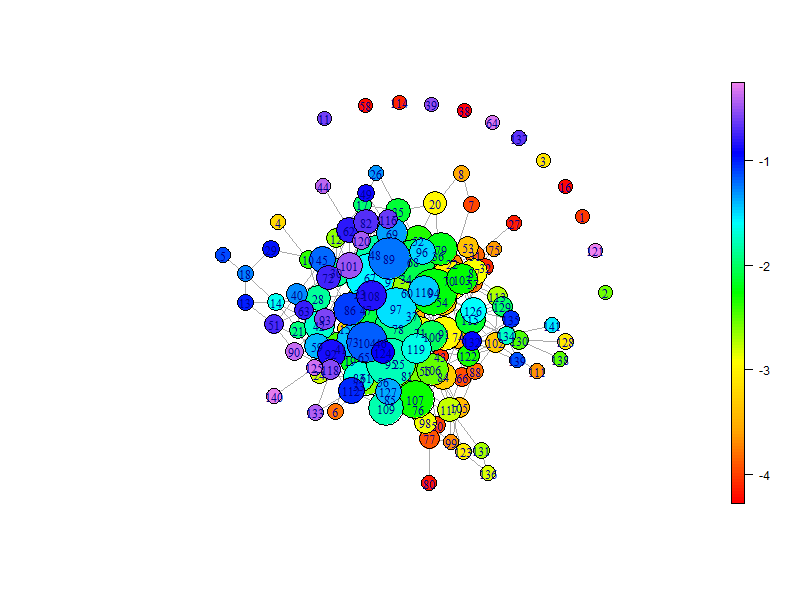}&\includegraphics[width=3.6cm]{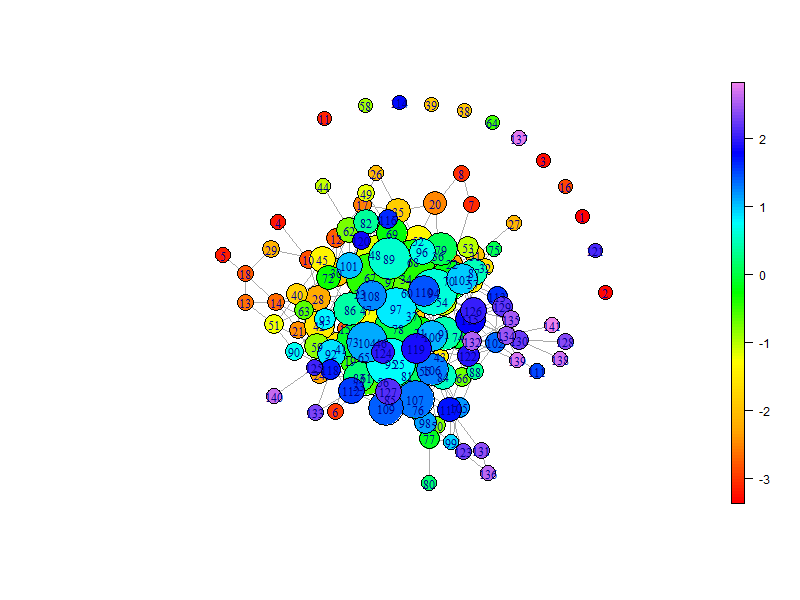}
			&\includegraphics[width=3.6cm]{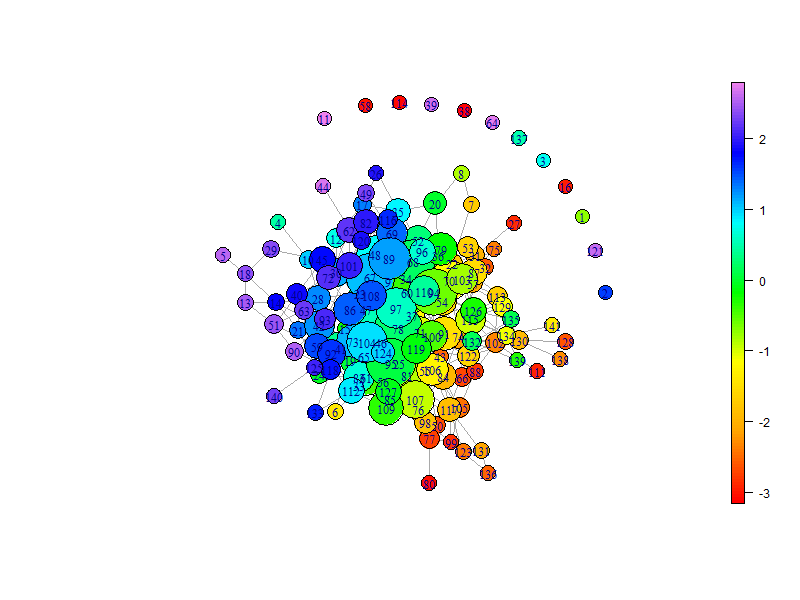}&\includegraphics[width=3.6cm]{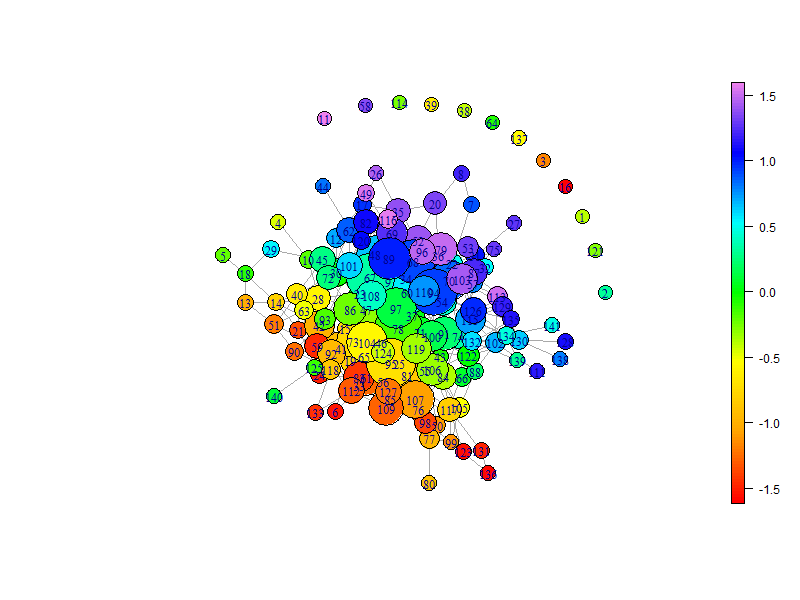}\\
			(e) Outcomes & (f) $x_0$ & (g) $y_0$ & (h) $z_0$ \\
			\includegraphics[width=3.6cm]{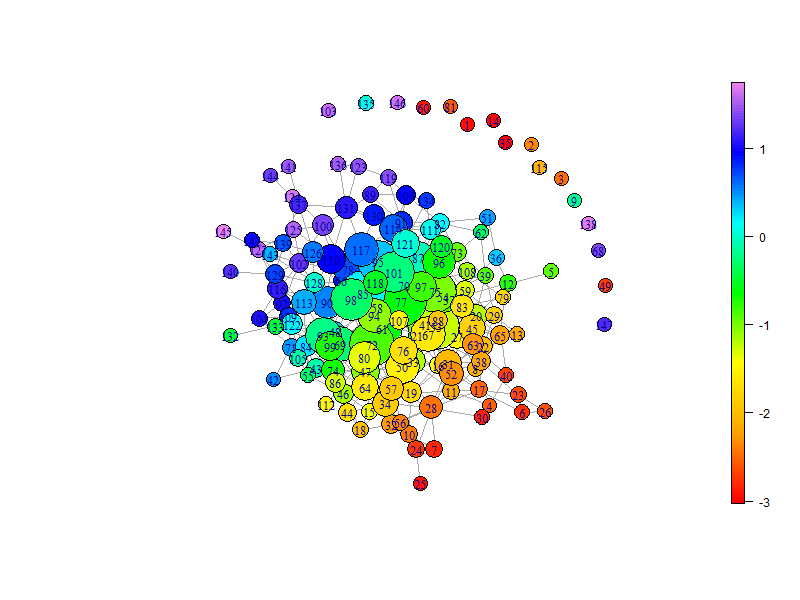}&\includegraphics[width=3.6cm]{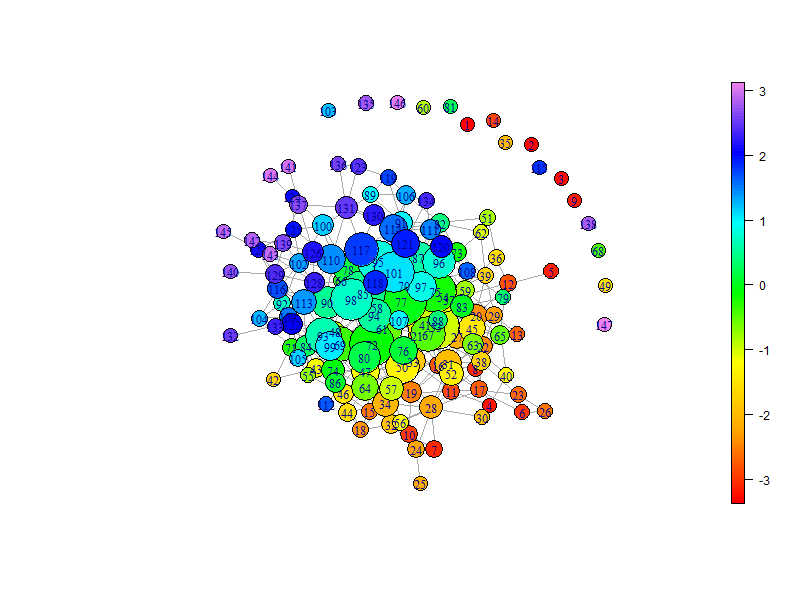}
			&\includegraphics[width=3.6cm]{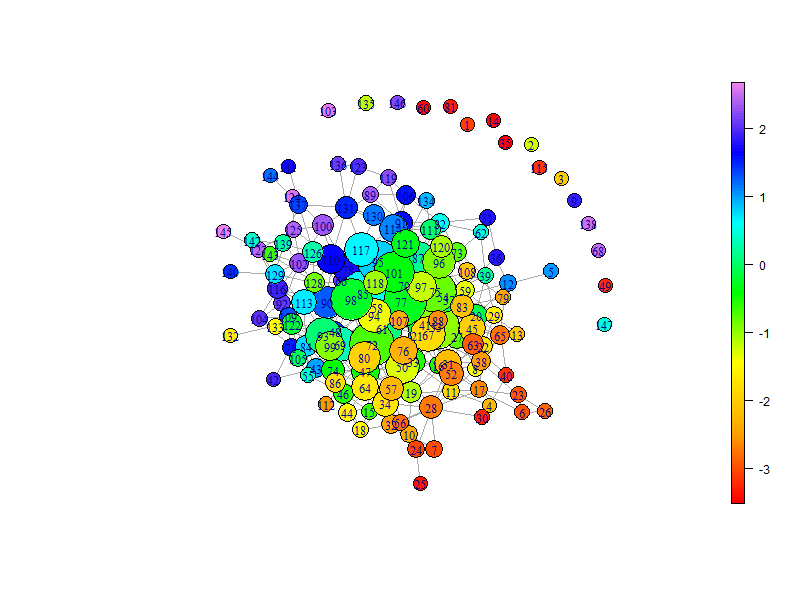}&\includegraphics[width=3.6cm]{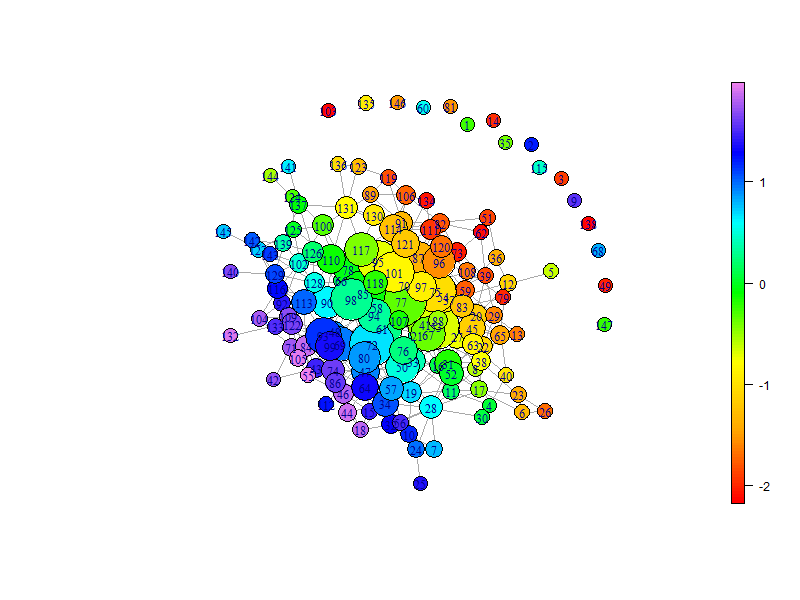}\\
			(i) Outcomes & (j) $x_0$ & (k) $y_0$ & (l) $z_0$ \\
		\end{tabular}
	\end{center}
\raggedright
\footnotesize{Notes: Left column coloured according to the outcome function $M=0.3x_0+0.6y_0+\omega$, where $\omega$ is an iid error term distributed $N(0,1)$. Column 2 colours by $x_0$, column 3 by $y_0$ and column 4 by $z_0$. Correlation matrices for row j are $\Sigma_j$ with $j=1,2,3$. $\Sigma_1=\begin{bmatrix}
	1 & -0.90 & 0.90 \\
	-0.90 & 1 & -0.62 \\
	0.90 & -0.62 & 1 \\
	\end{bmatrix}  $,
    $\Sigma_2=\begin{bmatrix}
    1 & -0.10 & 0.10 \\
    -0.10 & 1 & 0.98 \\
    0.10 & 0.98 & 1 \\
    \end{bmatrix}  $, and
     $\Sigma_2=\begin{bmatrix}
    1 & 0.40 & -0.40 \\
    0.40 & 1 & 0.68 \\
    -0.40 & 0.68 & 1 \\
    \end{bmatrix}  $. All plots produced using \textit{BallMapper} \citep{dlotkor} with filtration level $\epsilon=0.7$.
    }
\end{figure}

Motivation for the use of tools like TDA Ball Mapper comes from an ability to readily visualise data that would be otherwise impossible to quickly digest with standard summary techniques. Hence it is useful to consider the addition of further dimensions. In the simplest case to add a third axis. Outcomes in Figure \ref{fig:cor2} are given by the same linear combination of two of the variables, $M_i=0.3x_{0i}+0.6y_{0i}+\omega$, where $\omega$ is an iid error term distributed $N(0,1)$. Figure \ref{fig:cor2} has the outcome plot in the left column with the axes used as the colouration variables for columns 2, 3 and 4. In each plot of the graphs there are outliers, these unconnected points are not sufficiently similar to members of the big mass to be connected there to. Comparing the colouration the dominance of $y_0$ is evident, the highest values in panels (a), (e) and (i) map directly to the high values of $y_0$ as illustrated in panels (c), (g) and (k) respectively. Correlations between $y_0$ and $z_0$ vary but there is always overlap between the dark blue shading. 

Irrespective of the number of axis variables a two-dimensional plot is produced. All variables will be at their minimum in one part of the cloud, and then have their value increase monotonically across the cloud. This is exactly as seen in Figure \ref{fig:cor2}. As these are draws from a normal distribution it follows that the centre of the plot is congested and that there will be some, but not many, flares coming out of the main group. Colouring by axis variable can be informative as to which of the dimensions is highest in that flare. However discerning amongst the large mass exactly which variable is driving the observed outcomes is much more difficult. For this reason users are recommended to also consult the summary statistics of the units in each ball if specifics are sought.    

\subsection{Summary of Panel Analysis}

This section primarily focused on the way that correlation shapes the point cloud and, in so doing, alters the nature of the TDA Ball Mapper plot that results. As the number of dimensions increases so the individual correlations become masked in main graph but the intuition linking them to the colouration prevail. In its simplest form the TDA Ball Mapper plot is a dimensional reduction, but it is an abstract one and should not be taken to be anything more. In the following sections further analysis functionality is introduced which can use the full topology of the underlying point cloud. 

\section{Application 1: Risk, Return and Inequality}
\label{sec:er}

\cite{jorda2019global} document the production of an extensive macroeconomic dataset which covers 16 countries between the 1870s and 2017\footnote{The data that is produced is available to download from \url{http://www.macrohistory.net/data/} with the version used in this paper being the extended version produced specifically for \cite{jorda2019global}}. The authors identify four key findings which are then explored here using TDA. Firstly that risky returns ($r^{risky}$) defined as a weighted average of returns on residential real estate and equity, have had very similar real returns throughout. Secondly, \cite{jorda2019global} note that safe returns ($r^{safe}$, the equally weighted returns on bonds and bills, have been very volatile. In turn, the third result is that the risk premium ($r^{prem} = r^{risky}-r^{safe}$) is volatile because of the variability in the safe return. Finally, motivated by \cite{piketty2014capital} and others \cite{jorda2019global} study changes in inequality ($ineq$) through the difference in the return to wealth ($r^{wealth}$) and the growth of GDP ($g$); $r^{wealth}>g$ meaning those with wealth were seeing returns rise faster. For this reason inequality is defined as $ineq=r^{wealth}-g$ Such speedier wealth increases for the already wealthy create rising inequality. This example thus focuses on $r^{safe},r^{risky},r^{prem},r^{wealth}$,$g$ and $ineq$. 

TDA Ball Mapper offers the opportunity to visualise the evolution of these variables collectively and to say more of what lies behind the patterns discussed in the bivariate, or trivariate, comparisons of \cite{jorda2019global}. What follows is an exposition of the way in which this is realised, beginning with a typical distributional approach to summarising data. \cite{ng2013facts}. \cite{dosi2015fiscal} stresses the importance of continually reappraising the evolution to inform policy. TDA Ball Mapper is then used to show the time variation through the parameter space, revealing how some nations presently exhibit similarities to their Great Depression positions, how others are closer to where they were in the 1960s, whilst a third group continue to evolve on a new path. Motivated by this a second set of mapper plots examines how far present growth rates are from those experienced in the Great Depression. 

\subsection{Data}

\begin{table}
	\begin{center}
		\caption{Countries Included in Dataset \label{tab:natno}}
		\begin{tabular}{l c l c l c l c}
			\hline
			Country & Obs & Country & Obs & Country & Obs & Country & Obs\\
			\hline
			Australia & 112 & France & 131 & Netherlands & 101 & Sweden & 132 \\
			Belgium & 105 & Germany & 108 & Norway & 127 & Switzerland & 112\\
			Denmark & 134 & Italy & 78 & Portugal & 67 & United Kingdom & 111\\
			Finland & 92 & Japan & 67 & Spain & 108 & United States & 124\\
			\hline
			
		\end{tabular}
	\end{center}
	\raggedright
	\footnotesize{Notes: Figures report the number of observations for each country. Variation is due to missing data and removal of any year with growth above 50\% or below -50\%. Data from \cite{jorda2017rate}. $N=1710$}
\end{table}

The \cite{jorda2019global} data is annual and covers 16 countries producing a raw set of 2499 country-years. After removal of missing values the number of years of data for each country varies greatly. extreme values the numbers of observations from each country differs further. Table \ref{tab:natno} reports we have just 67 observations for Portugal and Japan, whilst there are 134 for Denmark. Our discussion of the great depression later in this example focuses only on those nations for whom data is available and hence excludes Italy, Japan and Portugal. In this paper a final set of 1710 country-years is used.

\begin{table}
	\begin{center}
		\caption{Summary Statistics (Percentages)}
		\label{tab:sumstat}
		\begin{tabular}{l l c c c c c c}
			\hline
			Variable &  & Mean & s.d. & Min & q25 & q75 & Max \\
			\hline
			Return on Safe Assets & $r^{safe}$ & 5.386 & 5.115 & -15.24 & 2.448 & 7.176 & 40.85 \\
			Return on Risky Assets & $r^{risky}$ & 11.16 & 10.99 & -23.85 & 5.204 & 16.10 & 128.2\\
			Risk Premium ($r^{risky}-r^{safe}$) & $r^{prem}$ & 5.774 & 11.53 &-40.75 & -0.306 & 11.38 & 129.2 \\
			Growth & $g$ & 6.779 & 7.581 & -30.71 & 2.748 & 10.01 & 49.29 \\
			Return on Wealth & $r^{wealth}$ & 9.996 & 9.044 & -23.21 & 4.998 & 13.88 & 114.4 \\
			Growth of Inequality ($r^{wealth}-g$) & $ineq$ & 3.217 & 9.477 & -39.05 & 1.503 & 7.683 & 109.0 \\
			\hline
		\end{tabular}
	\end{center}
\raggedright
\footnotesize{Notes: All figures expressed as percentages. Data from \cite{jorda2017rate}. Country-years with missing data, or growth rates above 50\% in absolute value are deleted. Full details of the calculations of returns are provided in \cite{jorda2019global} and the supporting documentation to the dataset. $N=1710$ }
\end{table}

\begin{table}
	\begin{center}
		\caption{Correlations}
		\label{tab:cor}
		\begin{tabular}{l c c c c c c}
			\hline
			& $r^{safe}$ & $r^{risky}$ & $r^{prem}$ & $g$ & $r^{wealth}$ & $ineq$ \\
			\hline
			Return on Safe Assets & 1 &&&&&\\ 
			Return on Risky Assets & 0.125 & 1 &&&&\\
			Risk Premium & -0.324 & 0.898 & 1 &&&\\
			Growth & 0.054 & 0.371 & 0.329 & 1 &&\\
			Wealth & 0.254 & 0.977 & 0.819 & 0.361 & 1 & \\
			Inequality & 0.199 & 0.636 & 0.518 & -0.456 & 0.666 & 1\\
			\hline
		\end{tabular}
	\end{center}
\raggedright
\footnotesize{Notes: Correlations calculated for pooled data from \cite{jorda2019global}.}
\end{table}

A typical approach to summary statistics is to consider the distribution of variables through the mean, standard deviation, minima and maxima of each variable. Secondly correlations are used as a first representation of association. Table \ref{tab:sumstat} presents such summary statistics for the six variables used to study the stylised facts from \cite{jorda2019global}. Volatility of the return on risky assets is higher, but it is the variation in $r^{safe}$ that is closest to it's mean. Growth percentages vary greatly, with a limit imposed at 50\% in absolute terms to remove the most notable of outliers. Even within this a large upper tail is seen. Returns on wealth, measured as the weighted average of returns on risky and safe assets, are notably higher than growth leading to rising inequality. A full discussion of these statistics and their interpretation is the purpose of the \cite{jorda2019global} paper. Correlations in Table \ref{tab:cor} reveal strong association between risky assets, the risk premium and the growth of wealth. Section \ref{sec:art2} demonstrated the role of correlation in TDA Ball Mapper plots showing how the cloud shape may be determined by the strength of association between the dimensions, even when there are more than two dimension axes being plotted.

\subsection{Evolutions}

Charting the development of each economy in the set each nation is treated individually in what follows, TDA Ball Mapper being applied to the set of characteristic variables set out in Table \ref{tab:sumstat}. Colour is set according to the average year of observation in the ball. Ball size is determined by the number of years within each ball. When working in multiple dimensions like this it is useful to understand where each characteristic is high/low and hence accompanying every plot is a set of six plots that are coloured according to each of the axes variables. For brevity this section reports only two examples. Following the discussion above the choice of filtration is a trade off between detail and simplicity; the aim being a parsimonious level of understanding. Consequently $\epsilon=0.04$ is used for the plots with results for other $\epsilon$ being available on request. 

\begin{figure}
	\begin{center}
		\caption{Evolution of Economic Returns with TDA Ball Mapper}
		\label{fig:evol}
			\begin{tabular}{c c c c}
				\\
				\multicolumn{2}{c}{SWEDEN} & \multicolumn{2}{c}{UNITED STATES OF AMERICA}\\
				\multicolumn{2}{c}{\includegraphics[width=7cm]{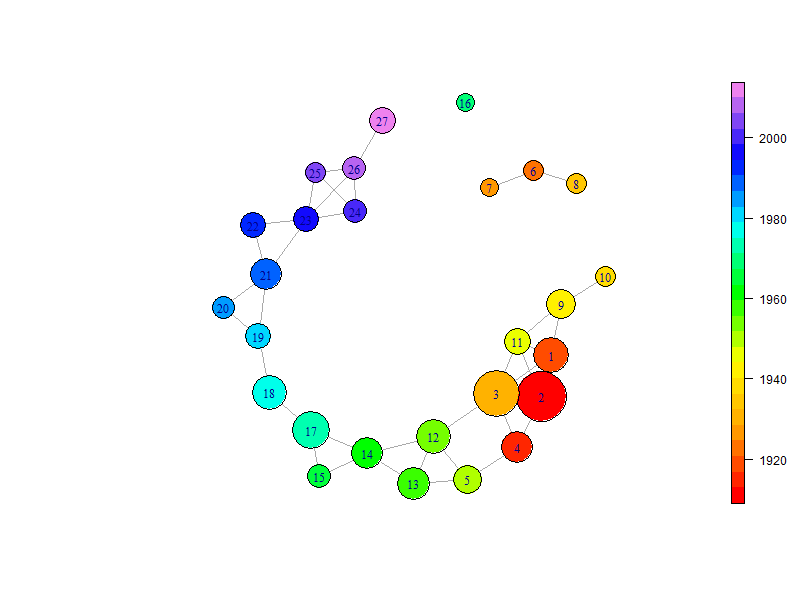}}&\multicolumn{2}{c}{\includegraphics[width=7cm]{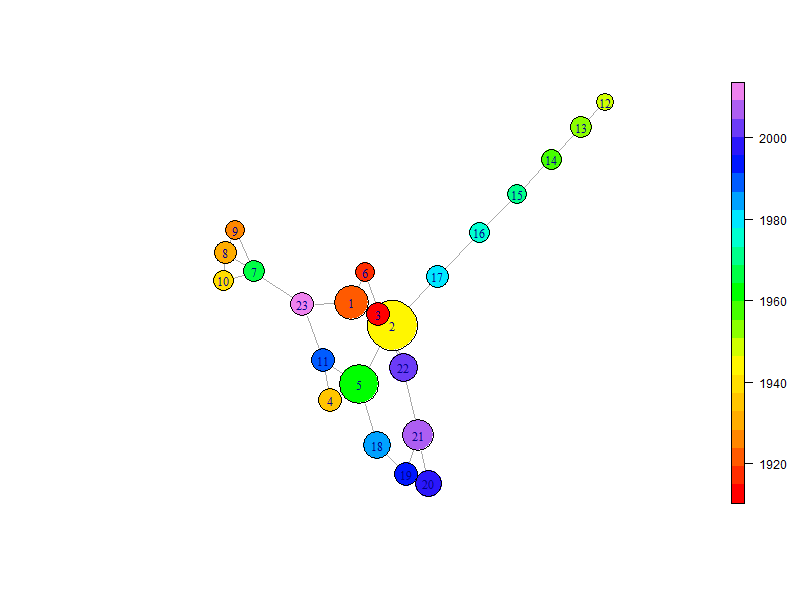}}\\
				\multicolumn{2}{c}{(a) Evolution} & \multicolumn{2}{c}{(b) Evolution} \\
				\includegraphics[width=3.6cm]{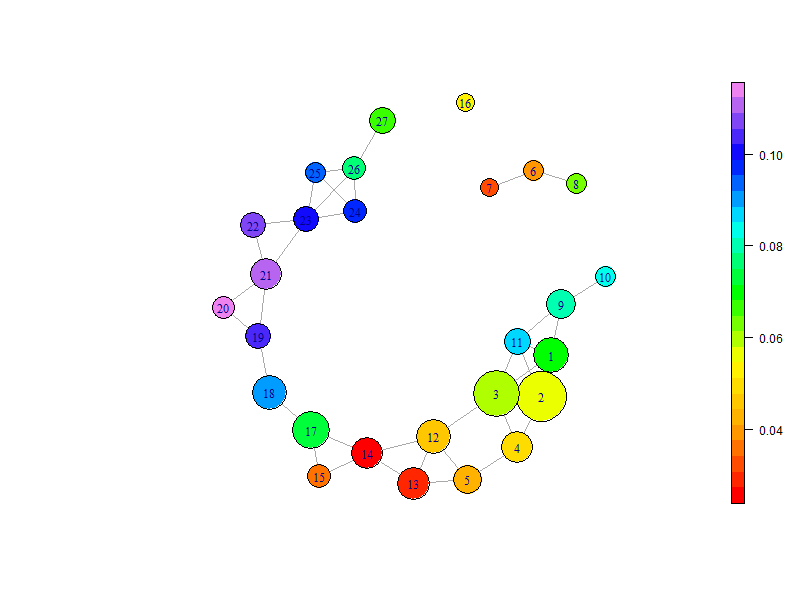} & \includegraphics[width=3.6cm]{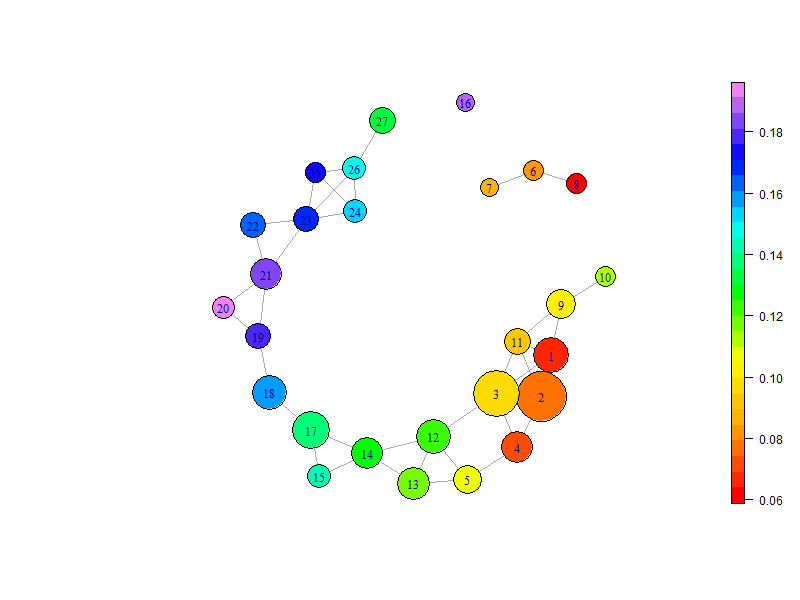} & \includegraphics[width=3.6cm]{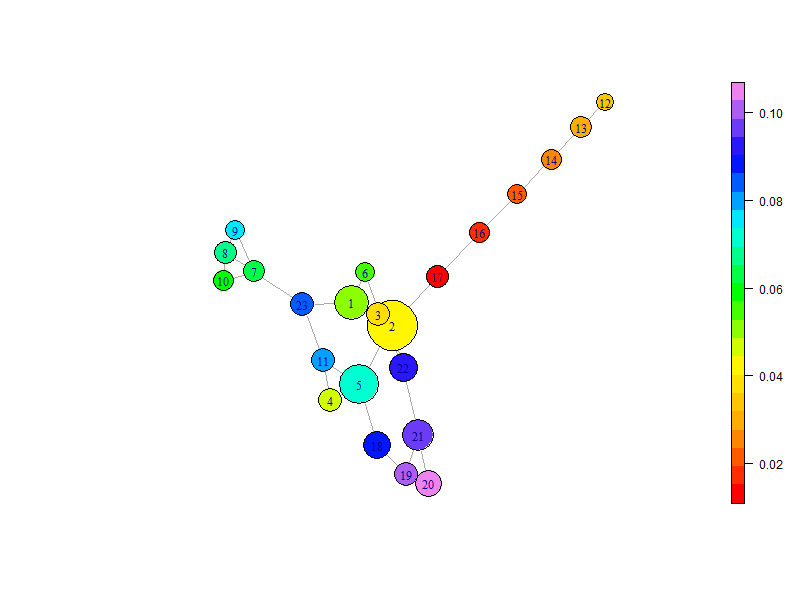} & \includegraphics[width=3.6cm]{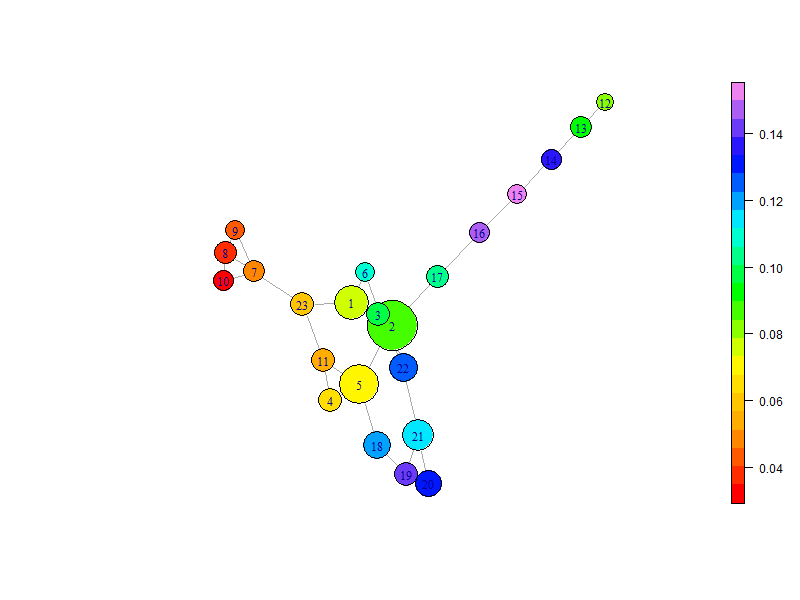}\\
				(c) $r^{safe}$ & (d) $r^{risky}$ & (e) $r^{safe}$ & (f) $r^{risky}$ \\ 
				\includegraphics[width=3.6cm]{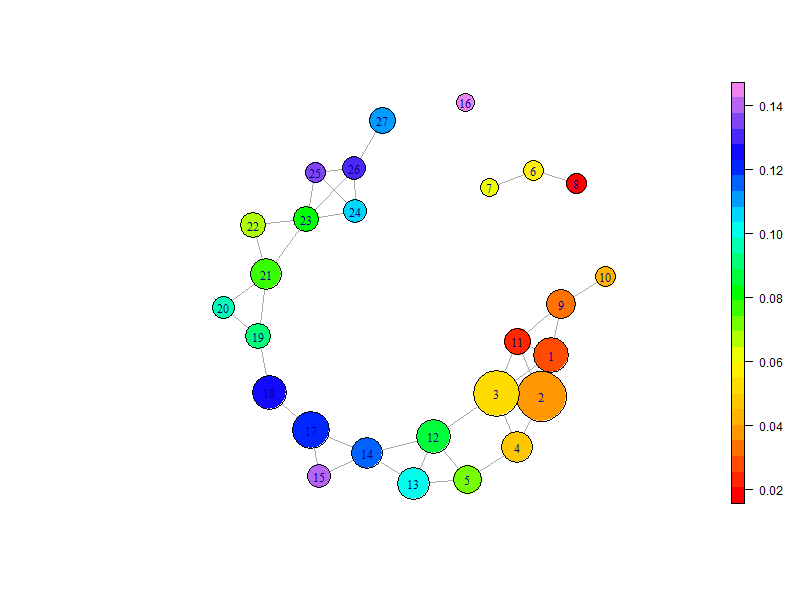} & \includegraphics[width=3.6cm]{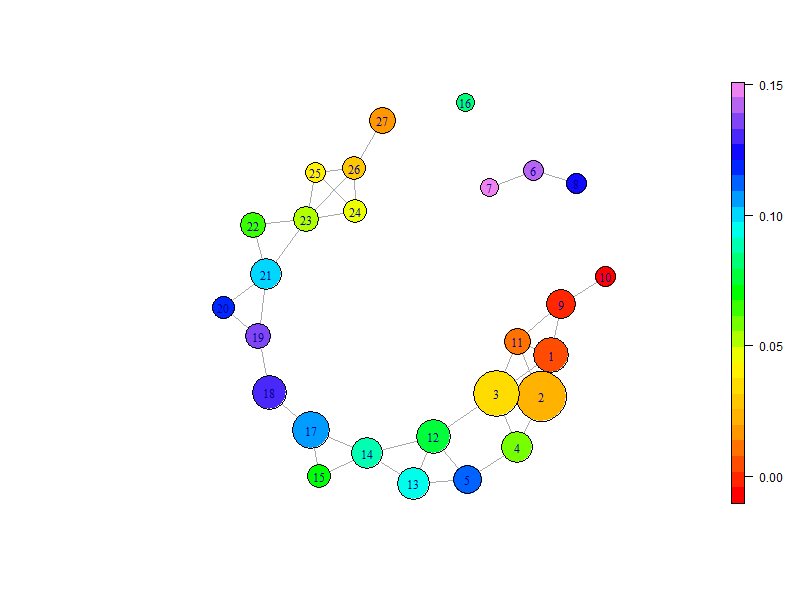} & \includegraphics[width=3.6cm]{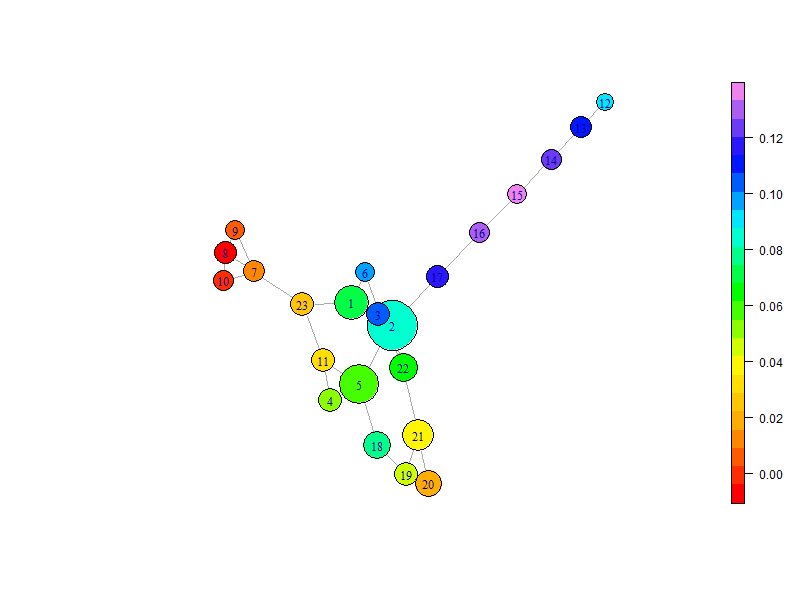} & \includegraphics[width=3.6cm]{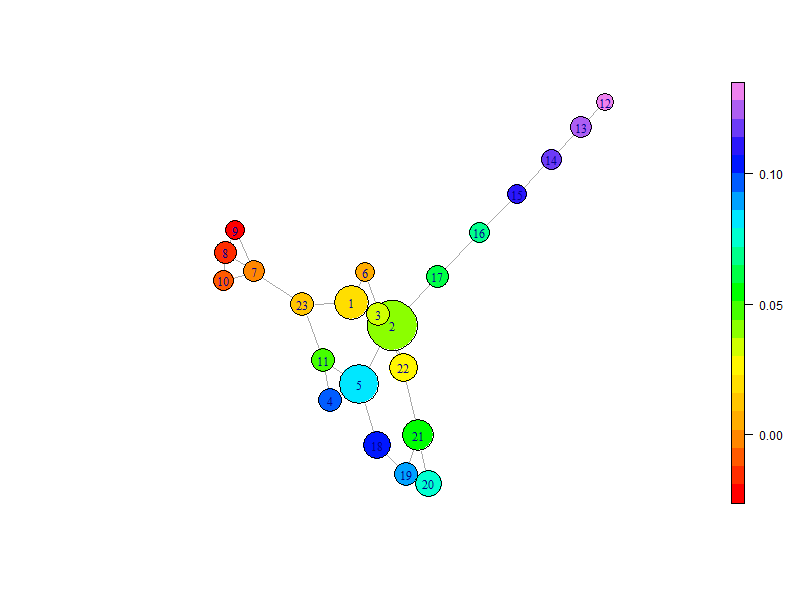}\\
				(g) $r^{prem}$ & (h) $g$ & (i) $r^{prem}$ & (j) $g$ \\
				\includegraphics[width=3.6cm]{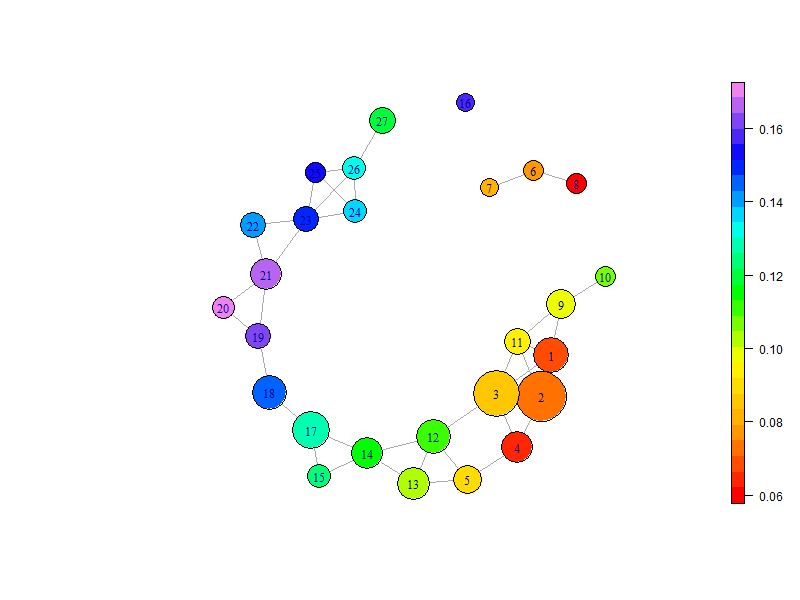} & \includegraphics[width=3.6cm]{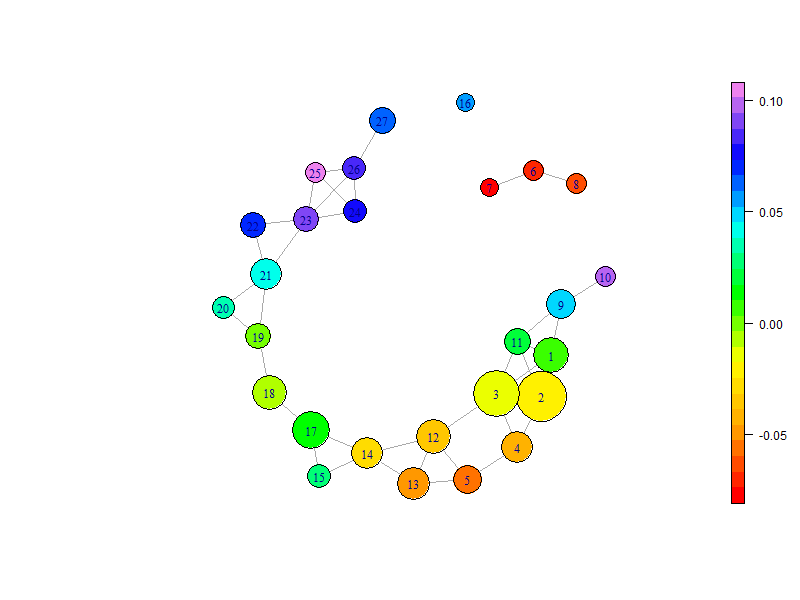} & \includegraphics[width=3.6cm]{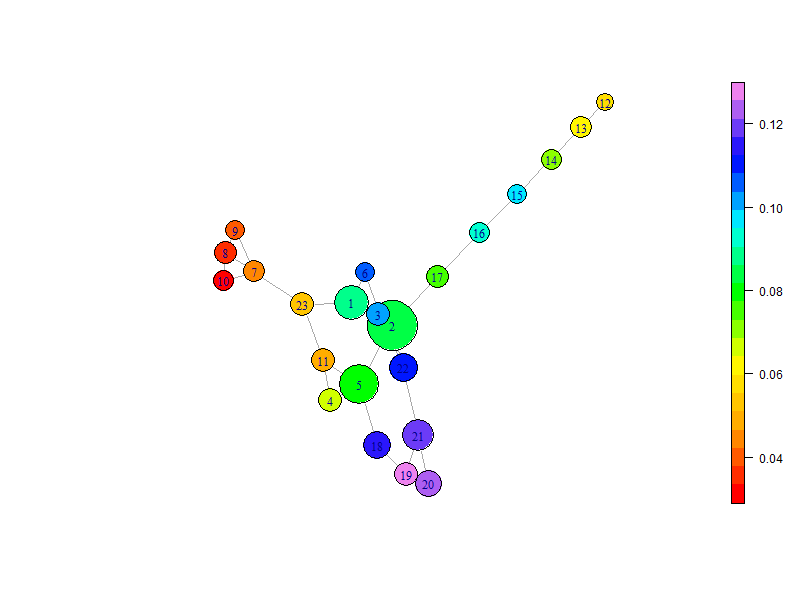} & \includegraphics[width=3.6cm]{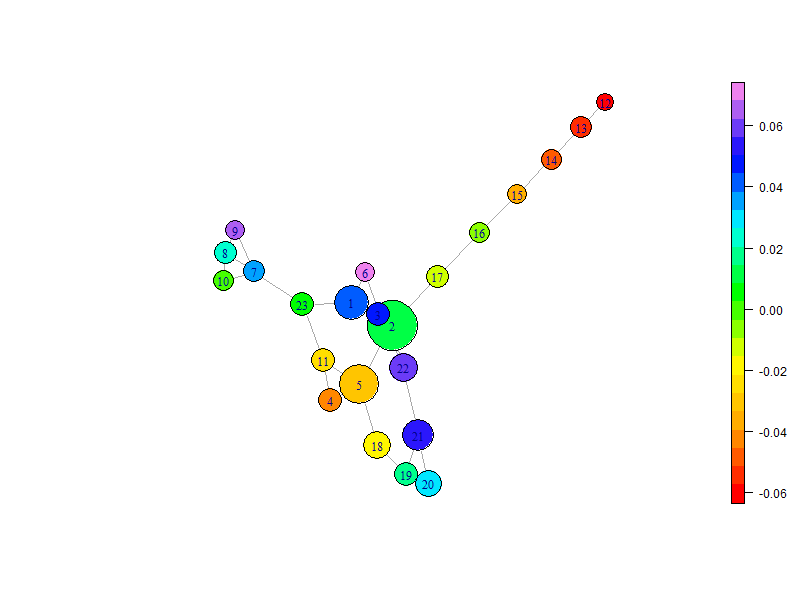}\\
				(k) $r^{wealth}$ & (l) $ineq$ & (m) $r^{wealth}$ & (n) $ineq$ \\ 
			\end{tabular}
	\end{center}
\raggedright
\footnotesize{Notes: Representative TDA Ball Mapper \citep{dlotkor} plots for Sweden and the United States of America. Main evolution plots, panels (a) and (b) are coloured by year. Lower plots are coloured according to the variable named below the plot. In all cases $\epsilon=0.04$ is used as filtration. Data from \cite{jorda2019global}.}
\end{figure} 

Figure \ref{fig:evol} plots two typical patterns that emerge from the study of the \cite{jorda2019global} data. On the left of the plot Sweden is an example of a country which has developed through time, rarely returning to historic values. Notwithstanding the outliers the plot charts a movement from the red balls of the 1920s (1,2 and 3) to the blues and pinks of contemporary days (24 to 27). There was a period immediately post World War 2 (9 10 and 11) that lay to the opposite side of the early balls to the rest of the economy, but this aside the path is clear. By contrast the USA, panel (b) shows a very different pattern. Having first been centred in the plot the Great Depression saw its position in space shift left before returning to the centre and ball 2. As the seat of the collapse in 1929 it is unsurprising that there is such variation shown by the data. Post World War 2 the economy was shocked greatly, heading up to ball 12 before charting a path back to the centre by the 1980s. More recently the USA has again displayed similar behaviour to the Great Depression, including a cycle that brought it back towards ball 2. Identifiable as the most contemporary ball, number 23 is back towards some of the leftmost balls observed in the Great Depression.

Such a variation is interesting but it is important to understand what this abstract representation actually means in terms of the characteristics. Figure \ref{fig:evol} helps with this. First consider Sweden. Returns on safe assets, panel (c) were at their lowest midway through the evolution, whilst returns on risky assets, panel (d), were at their lowest in those early years. Both reach their maximum in current times. Consequently the risk premium has changed a lot over time. Lowest values are seen in those post World War 2 balls on the right of the chain, with the maximal values coming where $r^{safe}$ is low but $r^{risky}$ has reached a middle value in the centre of the chain. More recently panel (g) also reports an increasing risk premium. Growth in Sweden, panel (h), loosely mirrors the risk premium, but the measure of wealth growth, $r^{wealth}$, shown in panel (k) is at its highest in the early years of the data. The very lowest growth of wealth came in the outlying set of three balls (6, 7 and 8) where it wealth was actually increasing at a negative rate. Panel (l) consequently is again mixed, but the negative wealth growth in balls 6, 7 and 8 mean that inequality was falling and hence there is redder colouration in panel (l). Bivariate plots like these are more akin to the comparisons undertaken in \cite{jorda2019global} but represent a valuable second stage in interpreting the main time variation plot.

Considering the USA those early balls showed mid-range values for both $r^{safe}$ in panel (e), and $r^risky$ in panel (f). It is the track back to the main cluster represented by the arm to the top right that is most interesting. Here $r^{safe}$ was falling but the risky return was achieving some of its highest values. Unsurprisingly the big negative returns are in the arm to the left of the plot and correspond to the height of the Great Depression. The contemporary cycle through balls 18 to 22 sees the safe return achieve some of its highest values despite the low interest rates. Likewise for $r^{risky}$, but here a fall is noted post Financial crisis. Risk premia in the USA were at their lowest during those Great Depression years of balls 8, 9 and 10. Growth, panel (j), shows clearly the cycle through recent years, obtaining a maximum mid way round the loop. On the trajectory back from the post war outlier growth also rises and falls in keeping with the economic cycle. In the TDA plot a loop isnt formed because none of the other variables connect ball 12 to any other ball. Panel (m) depicts the return to wealth as being at its' highest on the current cycle, but again there have been slight falls in recent years. Quicker falls in the growth rate mean that panel (n) shows inequality to have continued to rise. 

\begin{table}
	\begin{center}
		\caption{Summary Statistics for TDA Ball Mapper Balls}
		\label{tab:balls}
		\begin{small}
		\begin{tabular}{l l c c c c c c c c c}
			\hline
			Country & Ball & $r^{safe}$ & $r^{risky}$ & $r^{prem}$ & $g$ & $r^{wealth}$ & $ineq$ & \multicolumn{2}{l}{Year} & $n$ \\
			& & & & & & & & Min & Max & \\
			\hline
			Sweden & 1 & 4.628&7.008&2.379&6.672&2.47&4.202&1892&1940&13\\
			&2&4.041&7.237&3.197&6.817&4.354&2.464&1895&1941&23\\
			&3&4.041&8.219&4.178&7.626&4.719&2.908&1898&2009&20\\
			&4&3.694&7.022&3.327&6.623&5.88&0.743&1899&1926&10\\
			&5&3.253&8.901&5.648&7.694&9.074&-1.379&1916&1955&8\\
			&6&3.138&7.468&4.33&7.019&13.116&-6.097&1918&1921&3\\
			&7&2.676&7.622&4.946&7.092&14.797&-7.705&1919&1920&2\\
			&8&4.338&6.149&1.81&5.994&10.826&-4.832&1921&1923&3\\
			&9&5.907&8.842&2.935&8.43&1.138&7.292&1928&1937&9\\
			&10&5.961&9.244&3.283&8.768&-0.726&9.493&1929&1931&3\\
			&11&6.125&8.106&1.98&7.832&2.528&5.304&1927&1939&7\\
			&12&3.365&10.898&7.532&9.103&7.75&1.353&1906&1953&12\\
			&13&2.643&10.684&8.041&8.597&8.751&-0.153&1946&1956&11\\
			&14&2.564&12.185&9.62&9.963&8.222&1.741&1944&1960&10\\
			&15&3.113&15.366&12.253&13.048&7.677&5.371&1960&1966&5\\
			&16&3.719&18.195&14.476&15.647&8.077&7.57&1964&1965&2\\
			&17&4.802&14.485&9.683&13.114&8.909&4.205&1960&1976&14\\
			&18&6.612&16.345&9.733&15.012&11.273&3.739&1968&1984&12\\
			&19&9.699&17.363&7.664&15.68&11.664&4.016&1982&1988&6\\
			&20&11.394&19.355&7.96&17.048&10.426&6.622&1988&1991&4\\
			&21&11.324&17.659&6.335&15.825&8.847&6.978&1986&1997&10\\
			&22&11.143&16.598&5.455&14.989&7.287&7.702&1993&1998&6\\
			&23&9.609&16.936&7.328&15.073&5.743&9.331&1996&2004&6\\
			&24&8.112&16.306&8.195&14.301&5.111&9.19&2000&2004&5\\
			&25&6.661&17.362&10.701&15.16&4.72&10.44&2004&2006&3\\
			&26&5.874&15.914&10.04&13.971&4.706&9.265&2003&2012&5\\
			&27&4.381&12.633&8.252&11.495&3.906&7.589&2008&2015&7\\
			USA & 1 & "3.551&9.206&5.654&8.837&4.038&4.798&1900&2014&16\\
			&2&3.21&9.314&6.105&8.503&5.71&2.793&1901&2013&29\\
			&3&3.087&11.114&8.027&10.748&5.663&5.085&1906&1927&7\\
			&4&3.461&8.08&4.619&7.486&9.673&-2.187&1918&1974&7\\
			&5&4.004&9.146&5.142&8.343&8.412&-0.069&1907&1978&20\\
			&6&3.692&11.719&8.027&10.939&3.818&7.121&1909&1929&3\\
			&7&3.927&5.459&1.533&5.093&0.555&4.538&1931&2015&5\\
			&8&3.982&3.257&-0.725&3.25&-0.791&4.041&1932&1939&6\\
			&9&4.034&4.017&-0.017&4.002&-2.269&6.271&1933&1935&3\\
			&10&3.849&3.207&-0.642&3.114&0.505&2.609&1937&1940&4\\
			&11&4.594&6.987&2.393&6.605&6.013&0.592&1941&2010&7\\
			&12&3.043&9.257&6.214&7.08&13.131&-6.05&1943&1945&2\\
			&13&2.463&11.1&8.637&7.416&12.157&-4.741&1944&1949&5\\
			&14&2.083&12.824&10.741&7.976&11.725&-3.749&1947&1950&4\\
			&15&1.605&15.25&13.645&9.452&10.32&-0.868&1950&1952&3\\
			&16&1.535&14.693&13.158&9.431&6.936&2.495&1952&1955&4\\
			&17&1.296&11.563&10.266&8.239&6.551&1.688&1955&1960&6\\
			&18&6.347&12.19&5.844&11.222&9.822&1.4&1925&1984&10\\
			&19&9.091&13.508&4.417&12.773&9.514&3.259&1982&1988&7\\
			&20&10.49&12.432&1.942&12.068&7.637&4.431&1986&1994&9\\
			&21&8.212&12.069&3.857&11.282&6.026&5.256&1988&2004&13\\
			&22&6.682&12.199&5.517&11.139&5.552&5.587&1927&2006&11\\
			&23&4.944&7.069&2.125&6.723&3.957&2.766&2008&2015&7\\
			\hline
		\end{tabular}
	\end{small}
	\end{center}
\raggedright{Notes: Summary statistics for balls generated using \textit{BallMapper} \citep{dlotkor} in R on data from \cite{jorda2019global}. All figures expressed as the average percentage return within each ball.} 
\end{table}

Supporting these stories we can obtain summaries of the various balls using the \cite{dlotkor} output. Table \ref{tab:balls} gives such a summary for Sweden and the USA. A look at the years reveals that they are not just consecutive, in cases the minimum and maximum year in a ball can be very different. Balls 1 and 2 in Sweden include observations from the start of the dataset as well as some from the 1940s. This informs us that Sweden actually returned back to its original position in the point cloud on its way to the growth path. For the higher ball numbers the years are roughly consecutive but do have an overlap in the way that connected balls require. For the USA the summary of balls 1 and 2 shows that the economy did pass through these two original balls on the way to the left edge. 

This application is designed to show how TDA Ball Mapper can provide a readily interpretable visualisation of a dataset. In so doing multivariate comparisons are facilitated and stories about the input data may be told. From this start point it would be possible to split the safe and risky returns into their constituent parts. As an exposition of the technique this analysis next turns to visualising the relationship of points to the economy of the Great Depression. A full inter-temporal exploration of the \cite{jorda2019global} data is left to future work.

\subsection{Relationship to the Great Depression}

Colouring with any TDA Ball Mapper plot can be based upon any variable, or function thereof. Previously average year within a ball was used but there are many occasions when distance within the plot would be of interest. Because of the abstract nature of the representation of multivariate space it does not follow that separation of points in the plot has any link to the true difference in characteristics of the points. In the evolution of the returns the USA showed clear cycles, whilst after a small cycle in the early years Sweden took a more monotonic path. Like many other countries data from the USA clearly identifies a similarity between present data and the Great Depression of the 1930s. This subsection duly re-draws the plots colouring by the distance between each ball and the Great Depression levels. 

\begin{figure}
	\begin{center}
		\caption{Distance from Great Depression Return Levels}
		\label{fig:gdep}
			\begin{tabular}{c c c c}
				\includegraphics[width=3.6cm]{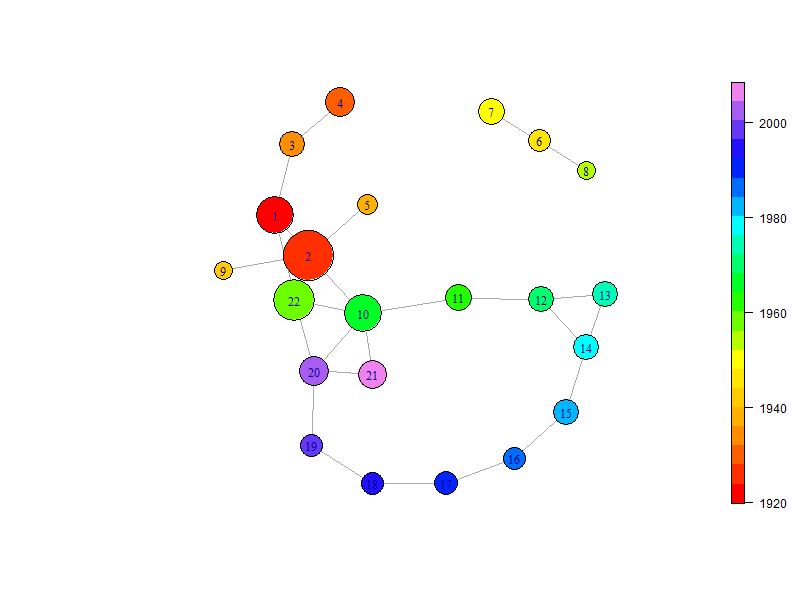} & \includegraphics[width=3.6cm]{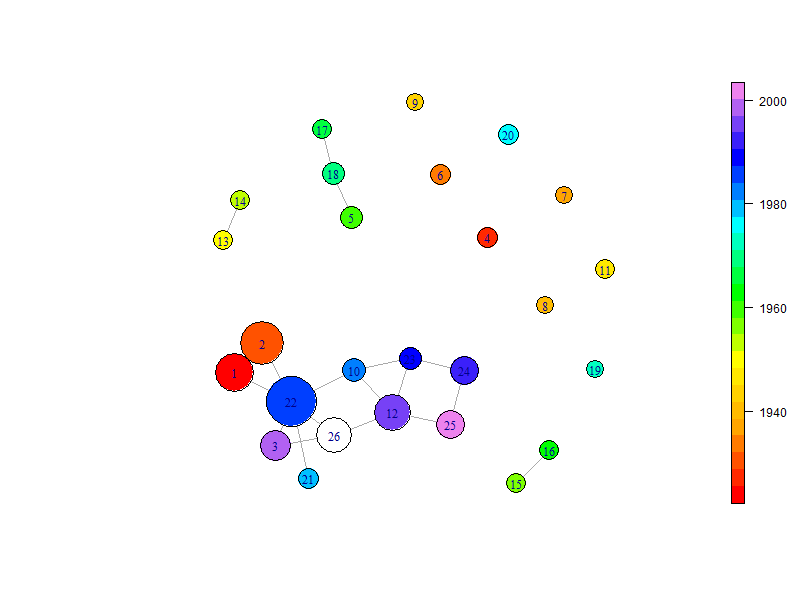} & \includegraphics[width=3.6cm]{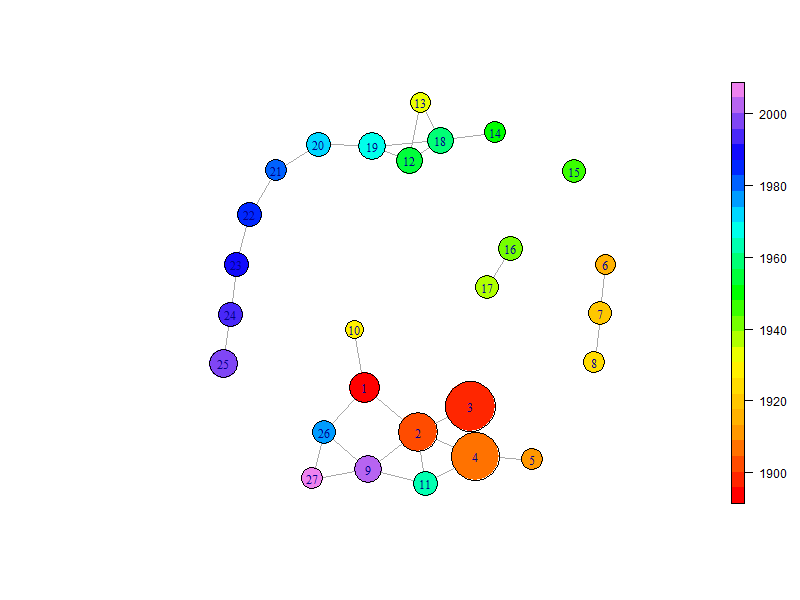} & \includegraphics[width=3.6cm]{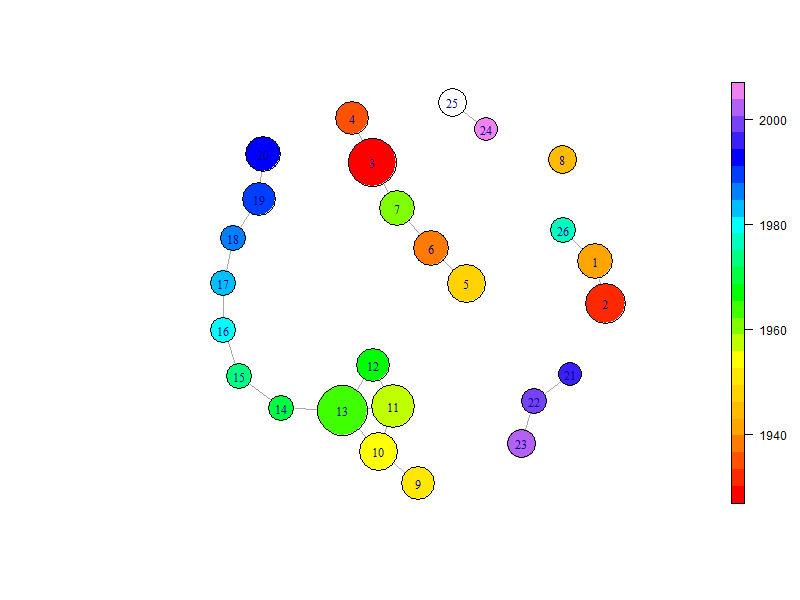}\\
				(a) Australia: Time & (b) Belgium: Time & (c) France: Time & (d) Spain: Time \\
				\includegraphics[width=3.6cm]{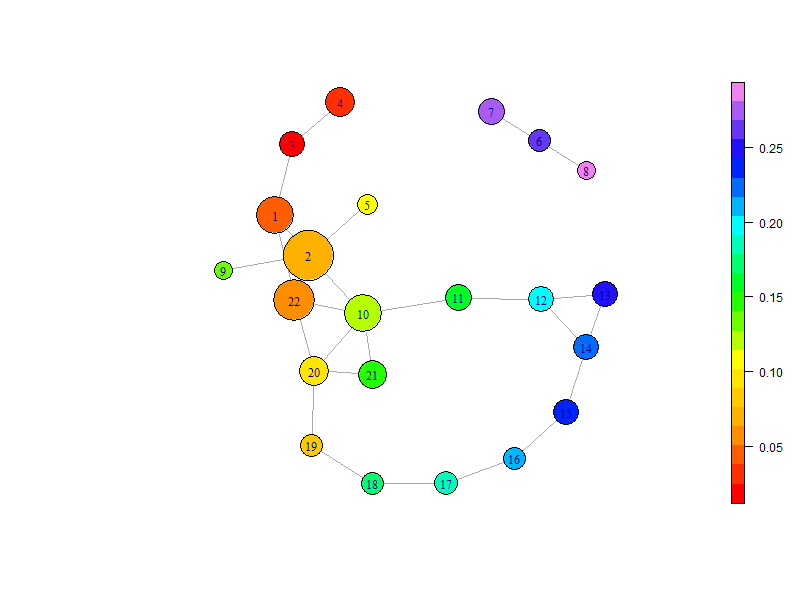} & \includegraphics[width=3.6cm]{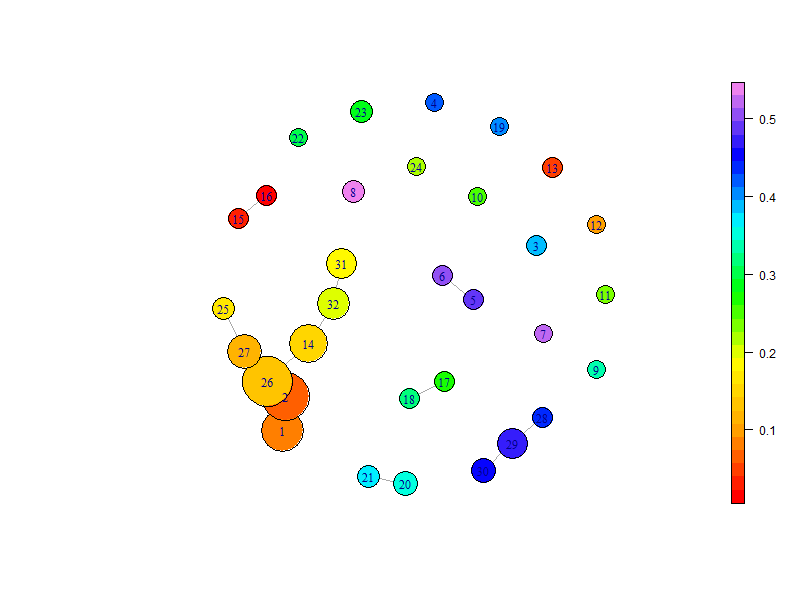} & \includegraphics[width=3.6cm]{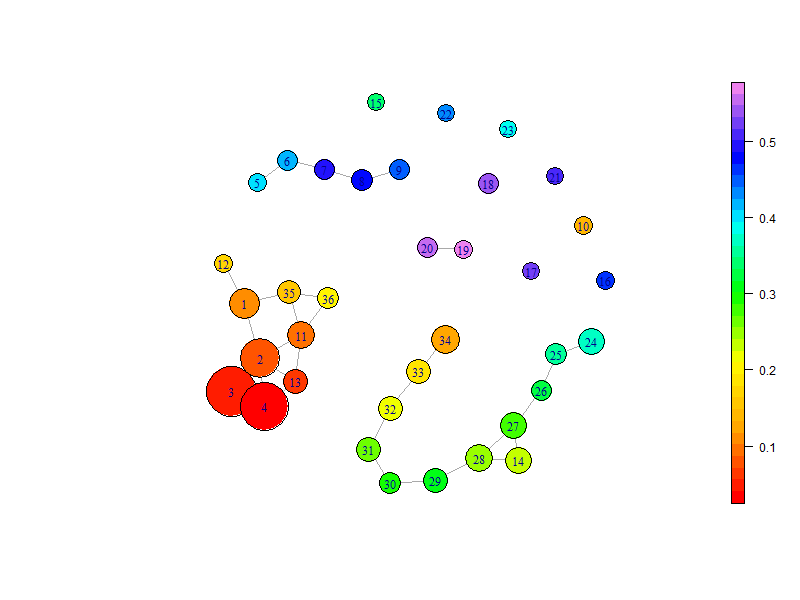} & \includegraphics[width=3.6cm]{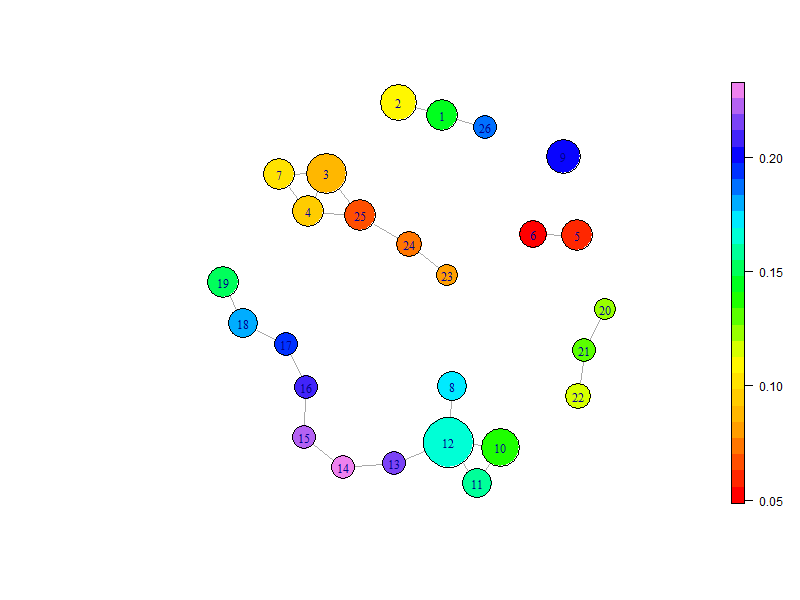}\\
				(e) Australia: Distance & (f) Belgium: Distance & (g) France: Distance & (h) Spain: Distance \\
				\includegraphics[width=3.6cm]{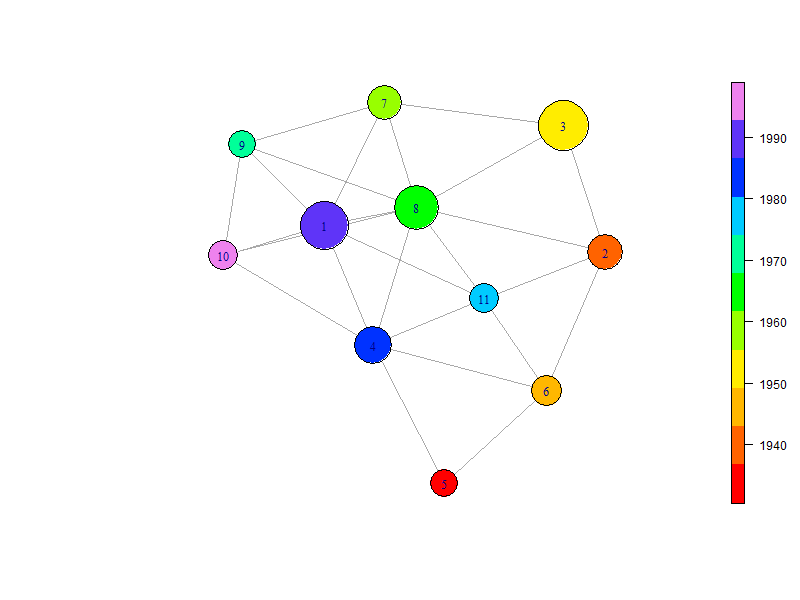} & \includegraphics[width=3.6cm]{SWEepx04.png} & \includegraphics[width=3.6cm]{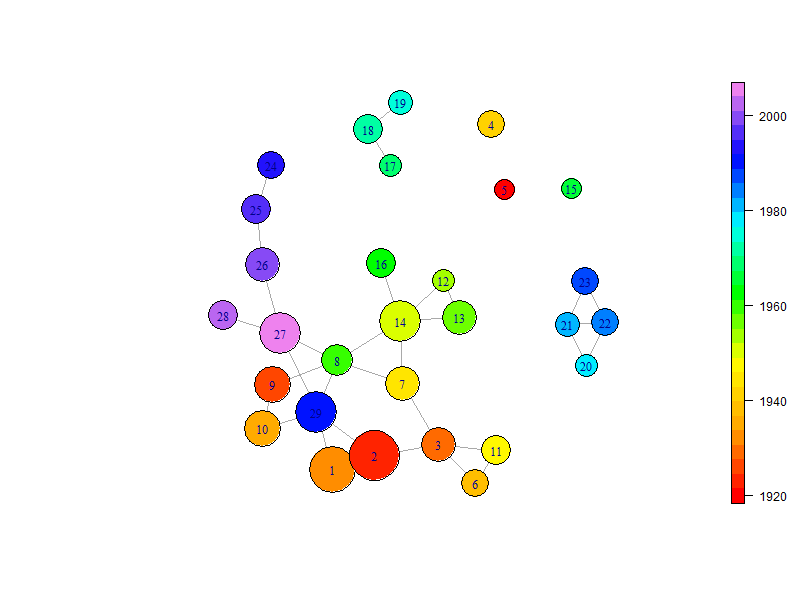} & \includegraphics[width=3.6cm]{USAepx04.png} \\
				(i) Switzerland: Time & (j) Sweden: Time & (k) UK: Time & (l) USA: Time \\
				\includegraphics[width=3.6cm]{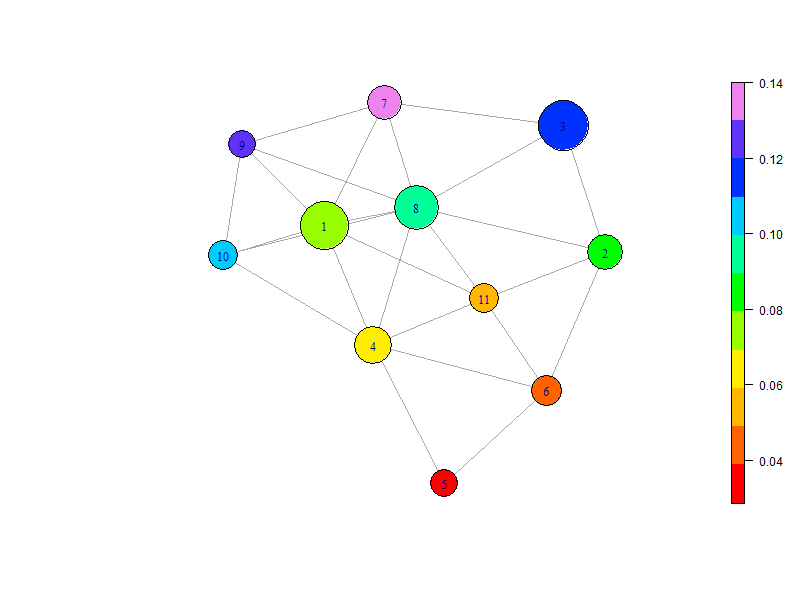} & \includegraphics[width=3.6cm]{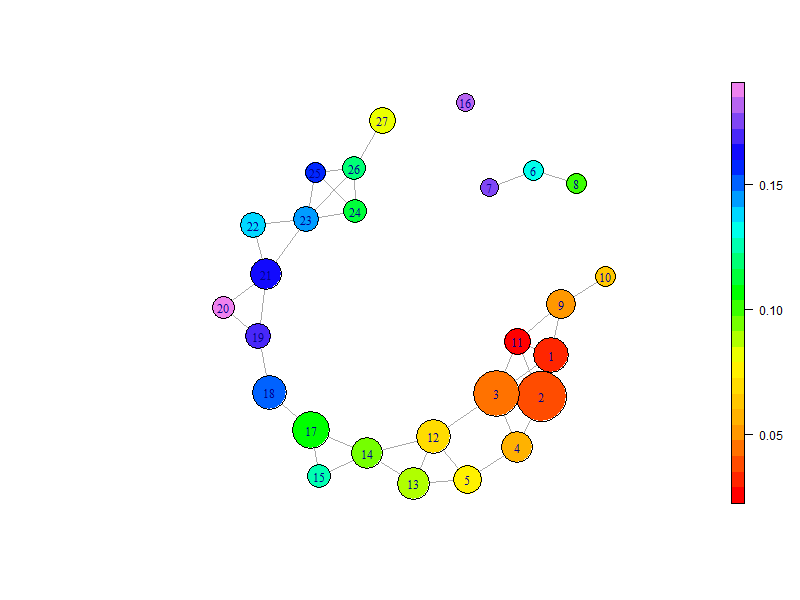} & \includegraphics[width=3.6cm]{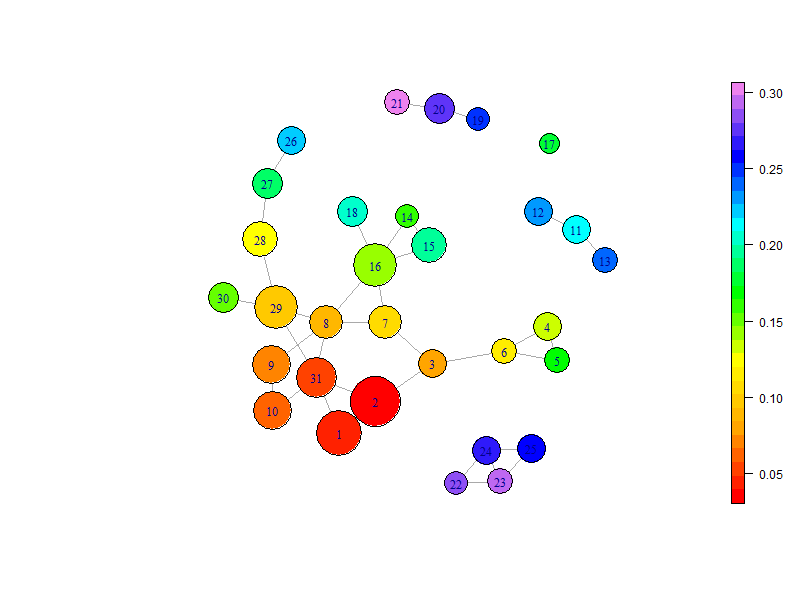} & \includegraphics[width=3.6cm]{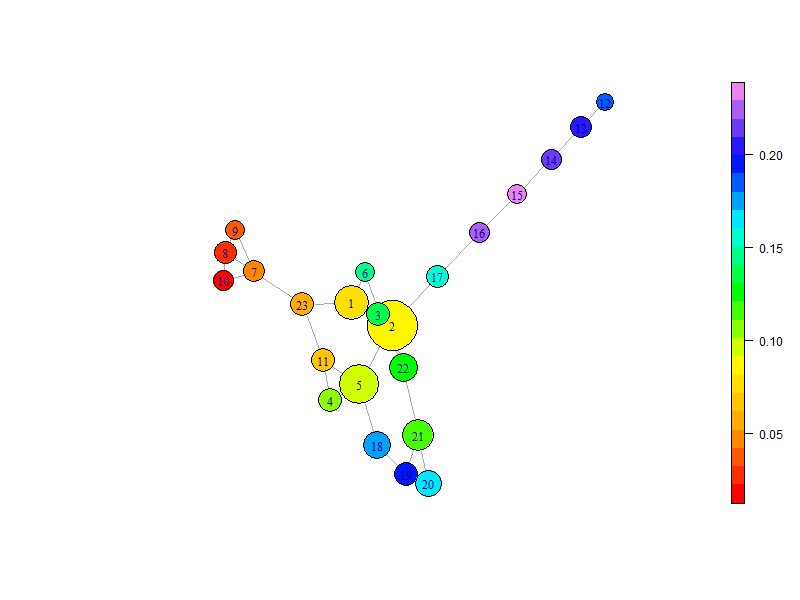} \\
				(m) Switzerland: Distance & (n) Sweden: Distance & (o) UK: Distance & (p) USA: Distance \\
			\end{tabular}
	\end{center}
\raggedright
\footnotesize{Notes: Plots show distance between the average axis value and the balls generated by the TDA Ball Mapper algorithm with $\epsilon=0.04$. Reference plots for Sweden and the USA can be found in Figure \ref{fig:evol}.}
\end{figure}

Studying the distance from the Great Depression consistently shows the high degree of similarity with contemporary observations. In many cases there are clear strings which bring the economy closer to the Great Depression levels. In the Australian plot, panels (a) and (e), the cycle over which the economy has travelled post war can be seen running round the bottom of the image getting further from the Great Depression average before coming back left towards the larger balls at the left centre. Belgian returns, panels (b) and (f), have been more volatile and hence there are fewer connected balls at $\epsilon=0.04$ than in most others. The abstract nature of TDA Ball Mapper plots is illustrated clearly by the seemingly random colouration of the disconnected outliers to the top and right of the image. France, panels (g) and (h) has a long string of points to the lower right, this is a typical evolution towards past values with the centre right end being the earliest observation in the string. Spain, in panels (d) and (h), has distinct groups of which one is closely linked to the Great Depression and the other is notably not. The temporal colouring for Spain shows that the long string getting closer to the depression levels, with its split end, is not the most recent observations. In fact the most recent returns in Spain are only mid way on the distance from the Great Depression scale. 

Switzerland in panels (i) and (m), is very different from the others as fewer balls are required to cover the data and there are no outliers. The connected nature of the graph means all points are close together but there is evidence that some of the more recent balls are close in characteristic to the returns Switzerland experienced during the Great Depression. Panels (j) and (n) show how the Swedish cycle pre World War 2 has all been coloured close to the Great Depression average, whilst the more contemporary points further away to the top left are also far from their historic levels. Panels (k) and (o) illustrate how the United Kingdom again finds itself again very close to its' Great Depression returns, the contemporary balls being in the deep orange region. The outlying cluster of four, not connected on any return variables, and far from the Great Depression values corresponds to the tough economic times of the late 1970s and early 1980s. Finally, in panels (l) and (p), the USA data displays the expected reversion to Great Depression returns. Particularly interesting in panel (p) is the way that the cycle travelled a long way from the Great Depression levels before it returned. The arm that sticks out towards the top right brings the data back towards the centre actually goes further from the depression before returning. Here again evidence of the economic cycle is seen.

\cite{fratianni2017tale} reviews this discussion of similarity in great detail, with particular emphasis on the comparative short time scale over which the global financial crisis has endured relative to the original Great Depression. \cite{fratianni2017tale} identifies similarities between the two big crises in that both were born of major asset crashes, equity for the Great Depression and housing for the global financial crisis, both were preceded by declines in international cooperation, and each was strongly linked to failures in the banking sector. This exposition did not segregate the two major risky asset classes, but the \cite{jorda2019global} data does incorporate an equity/housing split that cound be brough t\cite{bianchi2019great} considers Markov-Switching models to detect when the economy goes back into the Great Depression regime; the insights from which suggest that the probability of being in a Great Depression state was higher in the 1980s than more recently. Indeed the set of variables that best define the 1930s do not suggest that 2008 should be seen as being in that depression regime. With a predicted probability of switching approaching 0.6 the global financial \cite{bianchi2019great} does place it as the most similar. Conclusions from these papers are consistent with the stories seen in the TDA; that the economies came close to replicating the point cloud of the Great Depression but that, as yet, they have not gone to the extremes observed in the 1930s.

\subsection{Summary of the Risk Return and Inequality Topology}

Using an extensive macroeconomic dataset this section has revisited the four stylised facts that are identified in the paper which details the data \citep{jorda2019global}. Perhaps unsurprisingly there were very different patterns in the evolution of the nations within the panel; some countries having displayed a continuous morphing of their economies whilst others were seen to return to the Great Depression levels in the way documented in \cite{fratianni2017tale}. TDA Ball Mapper enables these patterns to be readily visualised and, through distance functions, to be further embedded in the colouration. Plots constructed from just four variables, and two linear combinations thereof, embed all bivariate and trivariate comparisons making them an efficient way to talk about temporal change. 

There are many ways in which the discussion could be extended, not least in adding further dimensions. Exploring the break downs between equity and housing returns, for example. Numbers of observations are limited, so some care must be taken, but a large advantage of the TDA approach is it is not reliant on sample size in the way that distributional methodologies are. Potential for colouration has not fully been exploited as other reference points may be introduced. Given the heterogeneity across countries in the visualisations nation specific events could be considered. All extensions would represent useful extensions to the \cite{jorda2019global} evaluation of their rich dataset.

\section{Private Credit Growth and GDP}
\label{sec:tdays}

As a second brief example TDA Ball Mapper is applied to the analysis of the link between the growth of private credit and GDP growth. Again this section uses the \cite{jorda2019global} dataset. Based on a discussion by \cite{steinkamp2018systemic} and a literature after \cite{ranciere2008systemic} the key issue is whether financial liberalisation, manifest in the growth of credit to households and private businesses, is producing increased growth. Of specific interest is whether negative skewness in the growth of real private credit is linked to economic growth. Should it be so, and it is shown comprehensively by \cite{steinkamp2018systemic} to be so, then crises could be considered beneficial to growth. This section then takes the TDA Ball Mapper lens to that relationship.

\subsection{Data}

Three axis variables are used for the analysis of private credit, each based on the growth rate of private credit. Over a rolling period of 10 years the mean average, standard deviation and skewness of year-on-year private credit growth is calculated. Table \ref{tab:pgsum} gives summary statistics for the countries included in the dataset. In their paper \cite{steinkamp2018systemic} add controls for the level of financial regulation, intial gdp and initial level of schooling in the countries. Because TDA Ball Mapper graphs are created for each country individually there is no need to have such controls here. Should these enter into the point cloud they would do so as single values and hence not contribute anything to the formation of balls.

\begin{table}
	\begin{center}
		\caption{Summary Statistics for Private Credit Growth}
		\label{tab:pgsum}
		\begin{tabular}{l c c c c c c c c}
			\hline
			Country & \multicolumn{4}{l}{Private Credit Growth} & Growth & \multicolumn{2}{l}{Data Range} & Obs \\
			& Mean & s.d. & Skewness & Sk Min & & Min & Max & \\
			\hline
			AUS&0.08&0.07&0.37&52&0.07&1880&2016&135\\
			BEL&0.16&0.36&0.64&27&0.11&1895&2016&107\\
			CAN&0.07&0.07&0.09&57&0.07&1880&2016&137\\
			CHE&0.05&0.03&-0.1&79&0.05&1880&2016&137\\
			DNK&0.07&0.04&0.28&47&0.06&1880&2016&137\\
			ESP&0.27&0.54&0.33&50&0.11&1910&2016&97\\
			FIN&0.12&0.01&0.44&38&0.10&1880&2016&137\\
			FRA&0.22&0.44&0.42&40&0.14&1910&2016&100\\
			GBR&0.07&0.06&0.27&54&0.06&1890&2016&127\\
			ITA&0.12&0.09&0.34&45&0.11&1880&2016&137\\
			JPN&0.15&0.14&0.78&16&0.15&1885&2016&131\\
			NLD&0.10&0.11&0.17&50&0.09&1910&2016&95\\
			NOR&0.08&0.06&0.29&51&0.07&1880&2016&131\\
			PRT&0.29&0.65&0.20&56&0.11&1880&2016&121\\
			SWE&0.07&0.05&0.33&47&0.06&1881&2016&136\\
			USA&0.07&0.06&-0.09&65&0.06&1890&2016&127\\
			\hline
		\end{tabular}
	\end{center}
\raggedright
\footnotesize{Notes: Growth values expressed in decimal form. Germany is omitted due to data issues. Sk Min is count of periods for which a negative value of Skewness is estimated. Min reports the first year for which data is available, Max the most recent. Columns 2 to 5 are calculated using rolling windows of length 10, the figures reported being the average of the rolling estimates. Data from \cite{jorda2019global}.}
\end{table}

In the sample there is some variation between the rates of growth of private borrowing, with Portgual showing an average of 29\% compared to just 7\% in countries such as the USA, UK and Canada. The USA and Switzerland both have a negative skewness in their growth distributions but others all display positive. In every case it can be seen that the countries had periods of 10 years where the skewness of their growth distribution was negative, meaning there is interest in looking at all of the nations in the panel. It is noted that Germany is dropped from this analysis rather than consider ways to account for the inter-war period. 

\subsection{TDA Ball Mapper Results}

\begin{figure}
	\begin{center}
		\caption{Real Private Credit Growth and Economic Growth: Australia}
		\label{fig:credit1}
		\begin{tabular}{c c c c c c}
			\multicolumn{3}{c}{\includegraphics[width=6cm]{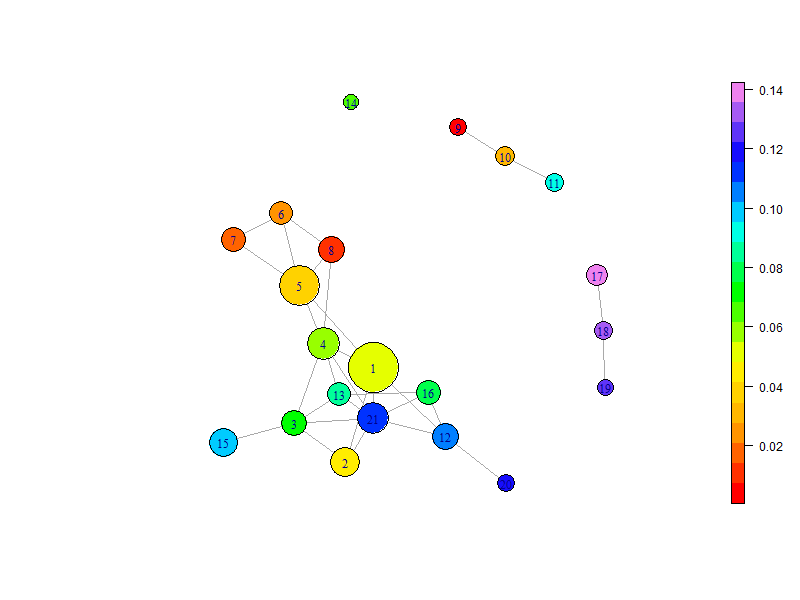}} & \multicolumn{3}{c}{\includegraphics[width=6cm]{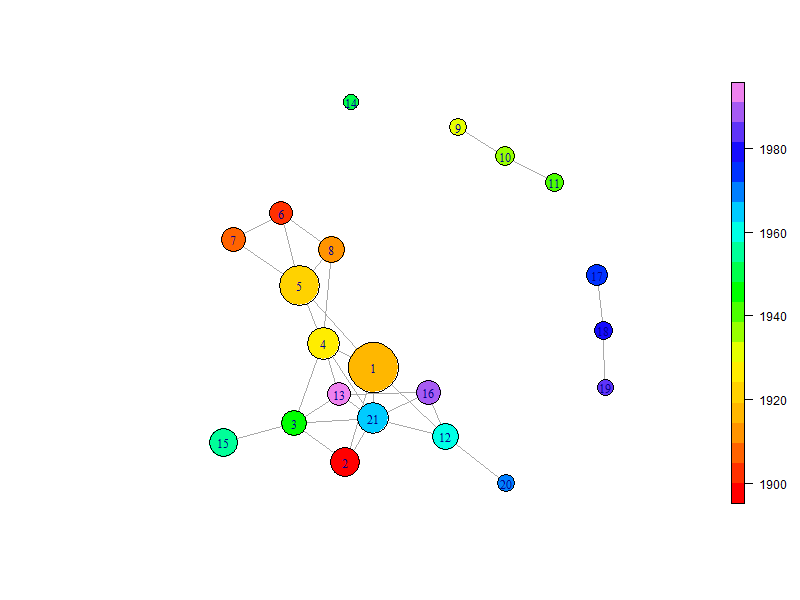}} \\
			\multicolumn{3}{c}{(a) GDP Growth} & \multicolumn{3}{c}{(b) Year}\\
			\multicolumn{2}{c}{\includegraphics[width=4cm]{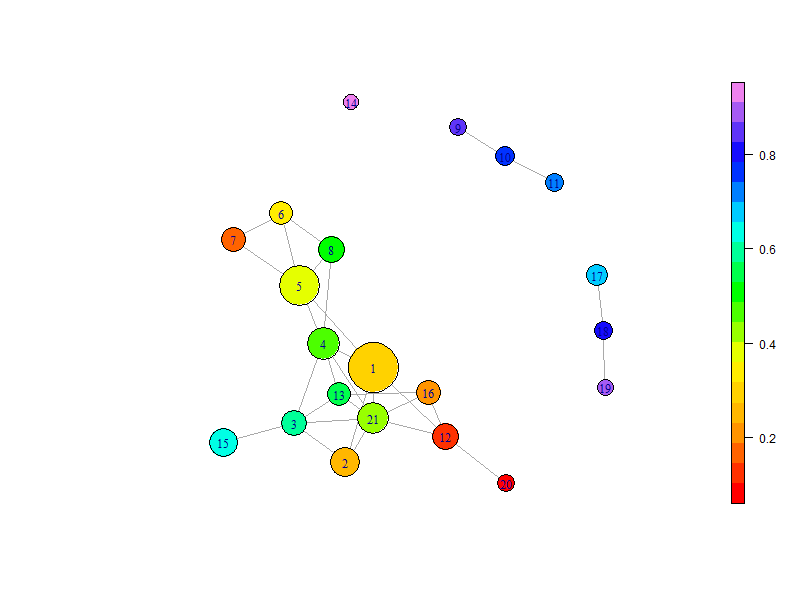}} & \multicolumn{2}{c}{\includegraphics[width=4cm]{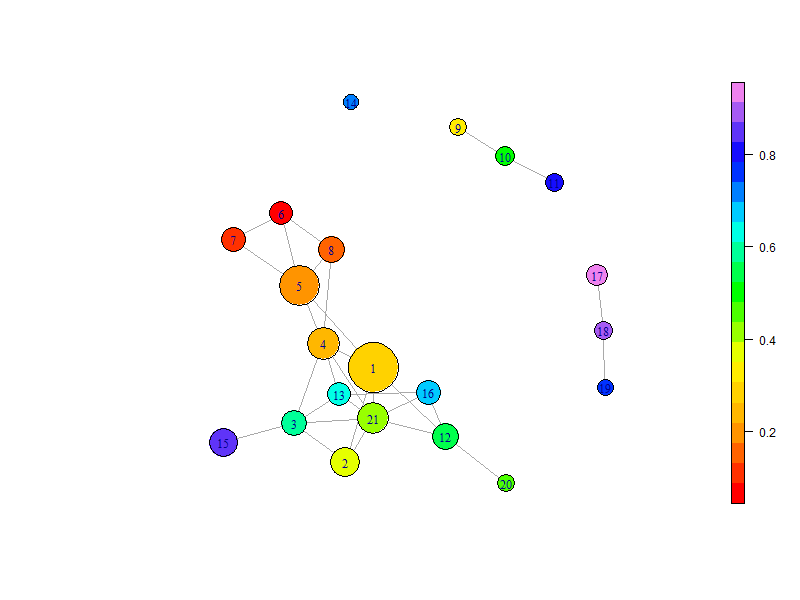}} & 
			\multicolumn{2}{c}{\includegraphics[width=4cm]{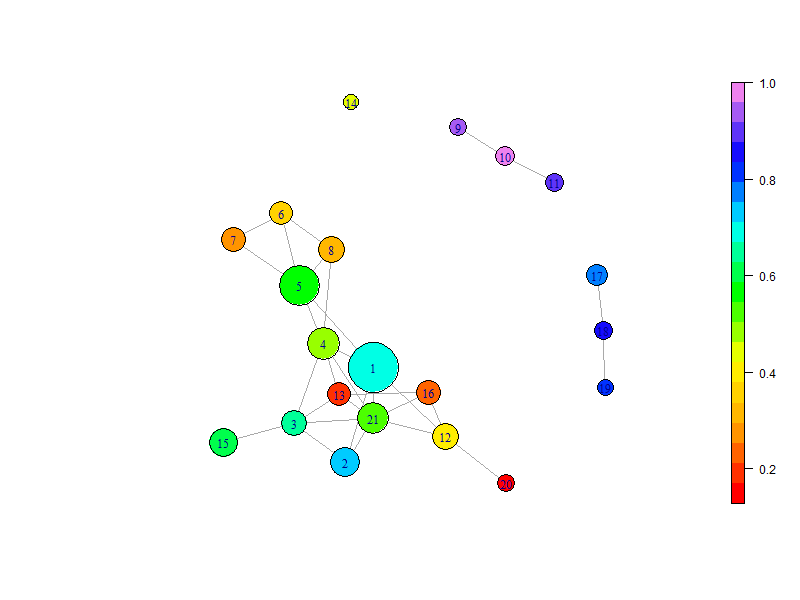}}\\
			\multicolumn{2}{c}{(c) Mean} & \multicolumn{2}{c}{(d) Standard deviation} & \multicolumn{2}{c}{(e) Skewness}
		\end{tabular}
	\end{center}
\raggedright
\footnotesize{Notes: Axis variables are the average, standard deviation and skewness of ten-year rolling windows of real private credit growth. Colouration in panel (a) is by the rate of economic growth, whilst panel (b) is coloured according to the year. The three lower panels (c) to (e) are coloured according to the axis variables. All TDA Ball Mapper plots generated using \textit{BallMapper} \citep{dlotkor}.}
\end{figure}

\begin{figure}
	\begin{center}
		\caption{Real Private Credit Growth and Economic Growth: United Kingdom}
		\label{fig:credit2}
		\begin{tabular}{c c c c c c}
			\multicolumn{3}{c}{\includegraphics[width=6cm]{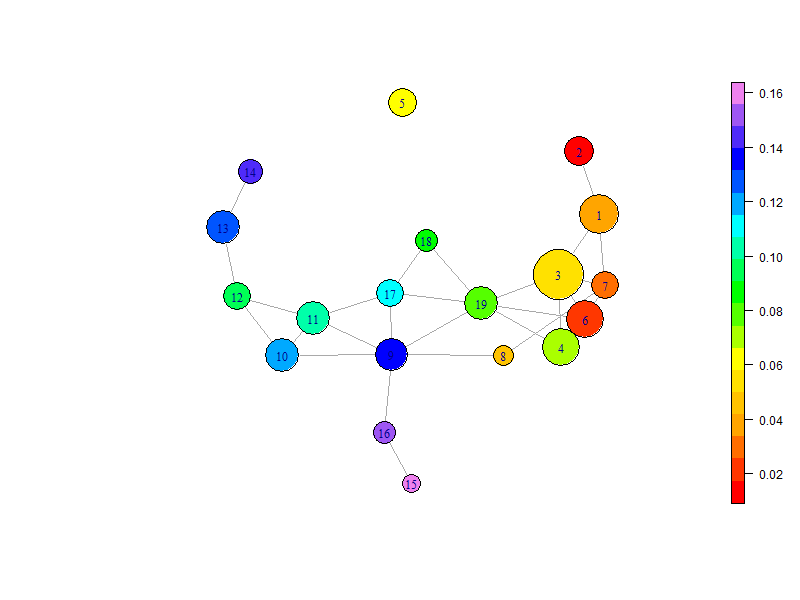}} & \multicolumn{3}{c}{\includegraphics[width=6cm]{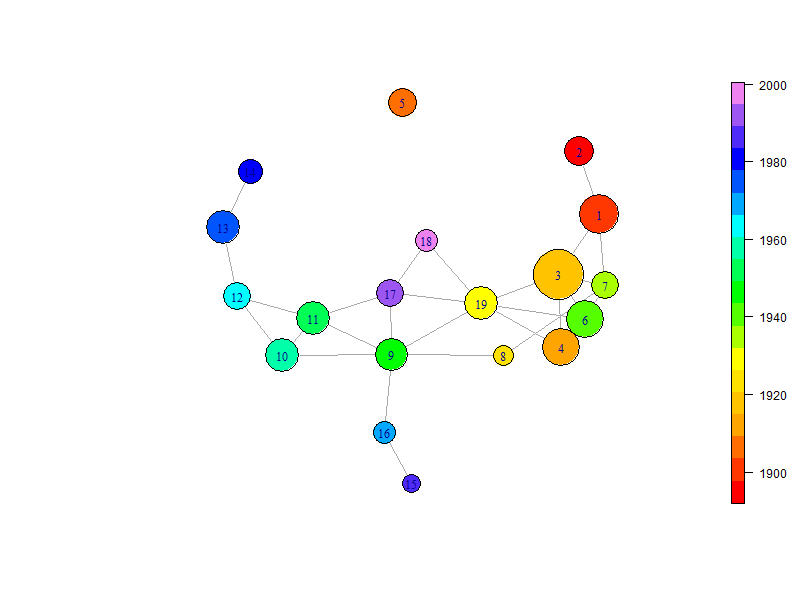}} \\
			\multicolumn{3}{c}{(a) GDP Growth} & \multicolumn{3}{c}{(b) Year}\\
			\multicolumn{2}{c}{\includegraphics[width=4cm]{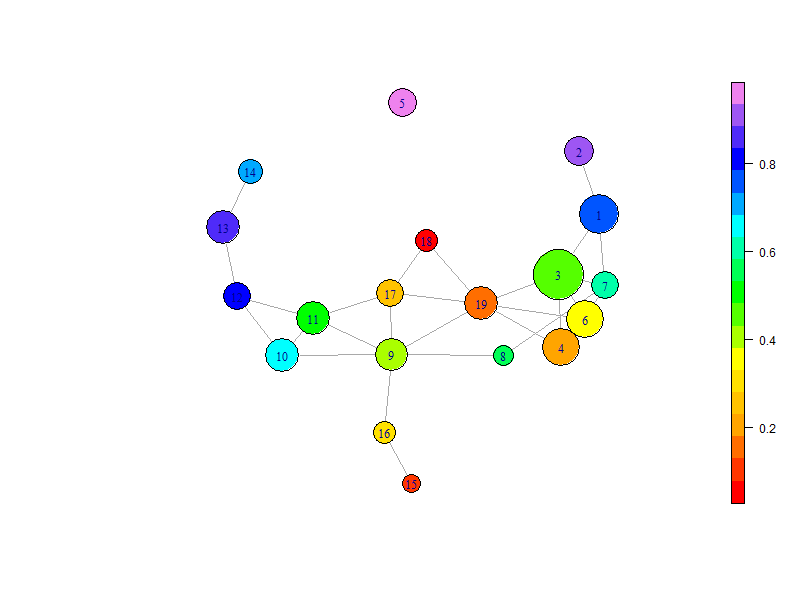}} & \multicolumn{2}{c}{\includegraphics[width=4cm]{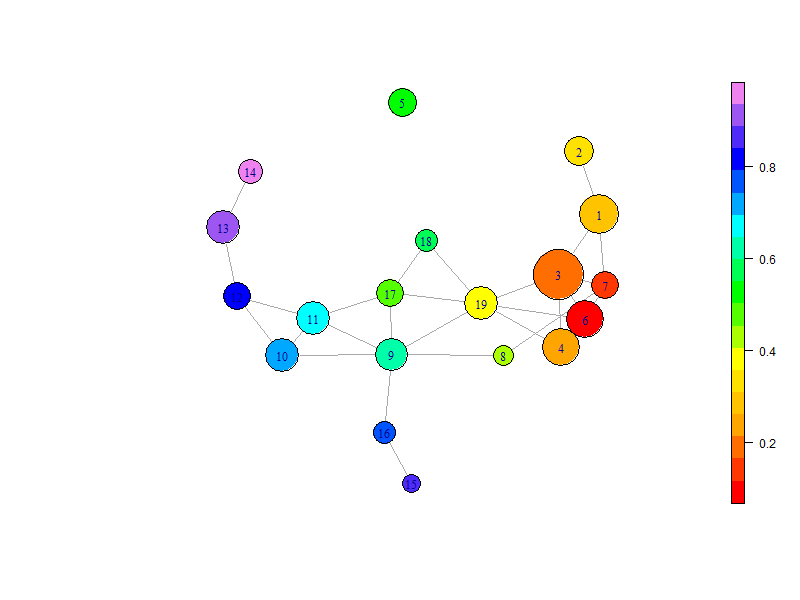}} & 
			\multicolumn{2}{c}{\includegraphics[width=4cm]{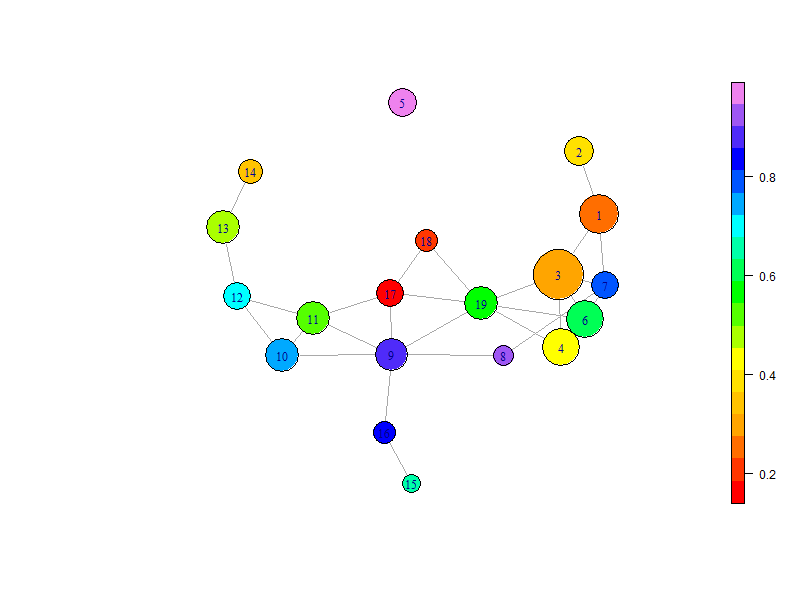}}\\
			\multicolumn{2}{c}{(c) Mean} & \multicolumn{2}{c}{(d) Standard deviation} & \multicolumn{2}{c}{(e) Skewness}
		\end{tabular}
	\end{center}
	\raggedright
	\footnotesize{Notes: Axis variables are the average, standard deviation and skewness of ten-year rolling windows of real private credit growth. Colouration in panel (a) is by the rate of economic growth, whilst panel (b) is coloured according to the year. The three lower panels (c) to (e) are coloured according to the axis variables. All TDA Ball Mapper plots generated using \textit{BallMapper} \citep{dlotkor}.}
\end{figure}

\begin{figure}
	\begin{center}
		\caption{Real Private Credit Growth and Economic Growth: United States}
		\label{fig:credit3}
		\begin{tabular}{c c c c c c}
			\multicolumn{3}{c}{\includegraphics[width=6cm]{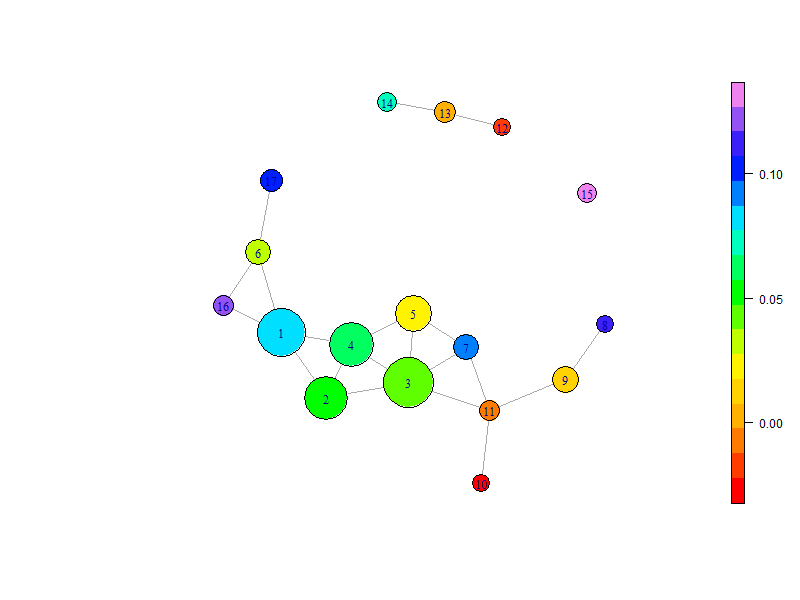}} & \multicolumn{3}{c}{\includegraphics[width=6cm]{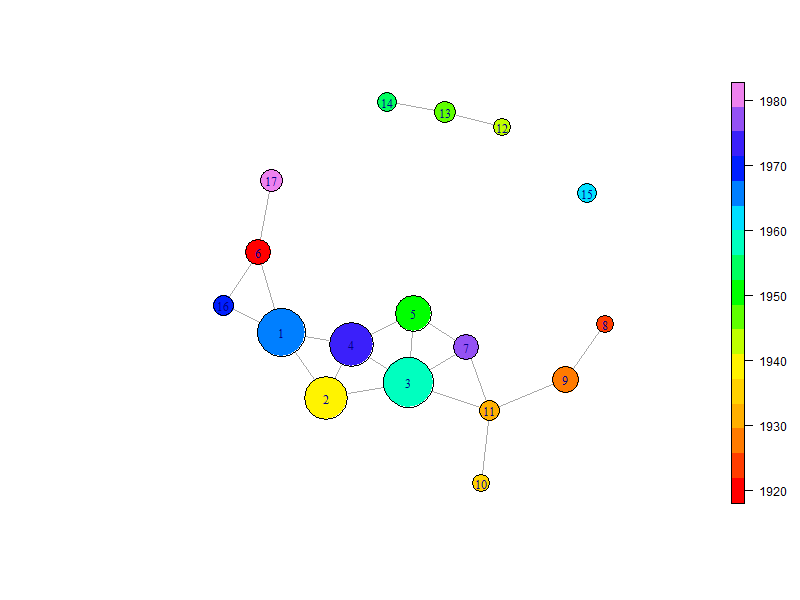}} \\
			\multicolumn{3}{c}{(a) GDP Growth} & \multicolumn{3}{c}{(b) Year}\\
			\multicolumn{2}{c}{\includegraphics[width=4cm]{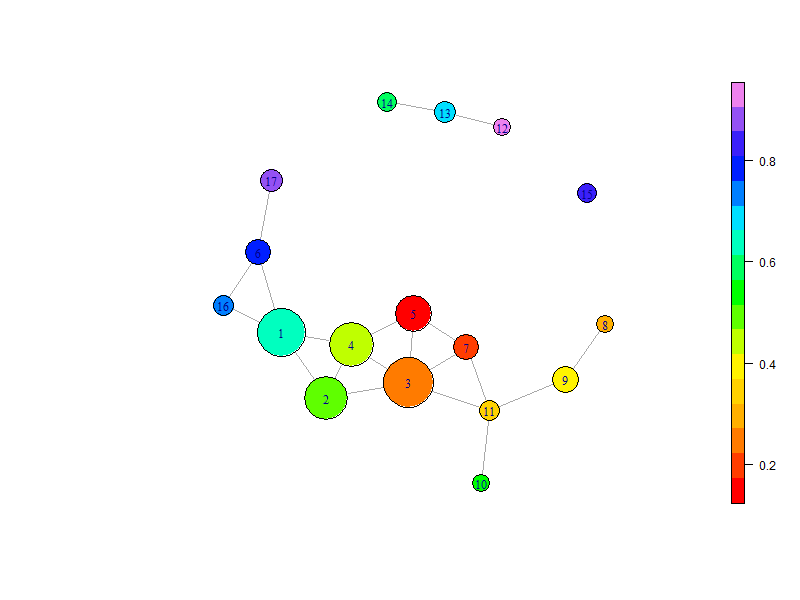}} & \multicolumn{2}{c}{\includegraphics[width=4cm]{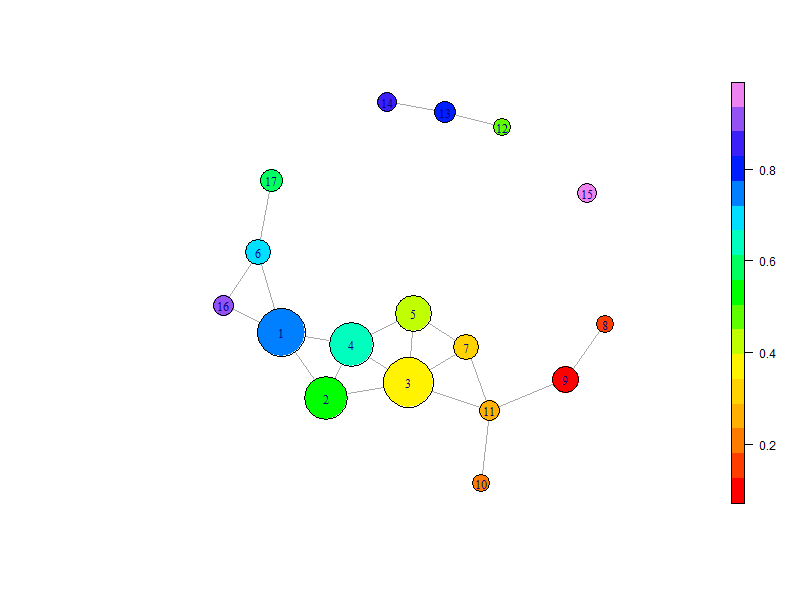}} & 
			\multicolumn{2}{c}{\includegraphics[width=4cm]{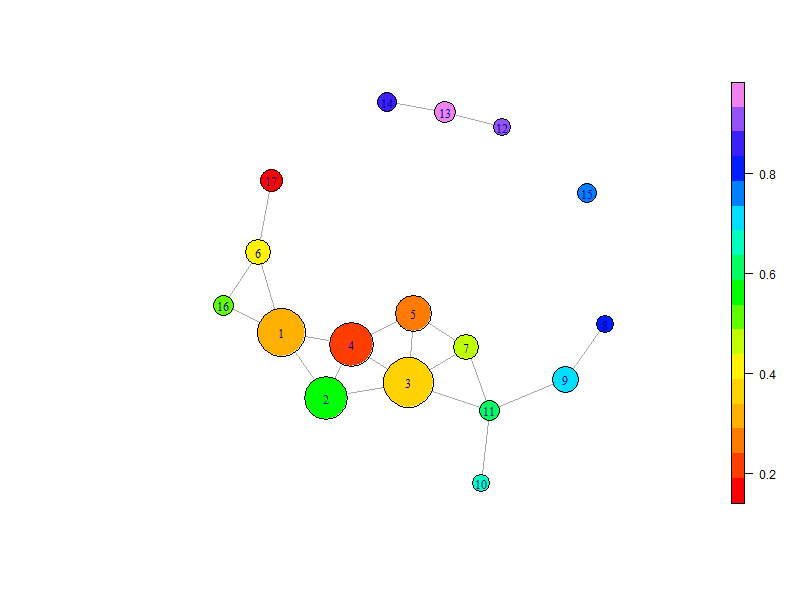}}\\
			\multicolumn{2}{c}{(c) Mean} & \multicolumn{2}{c}{(d) Standard deviation} & \multicolumn{2}{c}{(e) Skewness}
		\end{tabular}
	\end{center}
	\raggedright
	\footnotesize{Notes: Axis variables are the average, standard deviation and skewness of ten-year rolling windows of real private credit growth. Colouration in panel (a) is by the rate of economic growth, whilst panel (b) is coloured according to the year. The three lower panels (c) to (e) are coloured according to the axis variables. All TDA Ball Mapper plots generated using \textit{BallMapper} \citep{dlotkor}.}
\end{figure}

For some nations the rates of growth show such similarity that the number of balls is very low, but for most there are interesting patterns to be seen. Figures \ref{fig:credit1} to \ref{fig:credit3} present plots with all three axis variables normalised onto the range [0,1]. Normalisation becomes necessary because in this case they are not growth percentages and appear on very different scales. 

In the Australian case there are some clear areas of high growth on the arm to the lower right of the plot; but these sit very close to some much larger lower growth balls (such as ball number 1). Other high growth percentages are found in the string of three balls that sits outside the remainder of the cluster. There is a second string that heads out south west from the main group, this has very middle levels of growth on average. Lowest growth comes to the north of the cluster in balls 6 and 7. In panel (b) colouration by year reveals these low growth years to be in the earlier part of the sample. Association of these outcomes with the year is achieved by looking at panel (b). High growth in the right arm from the cluster has been more recent, as has the outlying string of three observations. Most recently Australia has found itself back near the centre of the plot with a combination of mean, standard deviation and skewness of its' real private credit growth combination closest to its most common (largest) ball. Colouring according to the three axis variables informs that the lower right arm from the cluster was indeed associated with negative skewness, and low average real private credit growth. However, the outlying group was not; in that string all three of the axis variables are close to their maximum.

As a second example consider the United Kingdom, plotted in Figure \ref{fig:credit2}. Rather like Australia there is a three arm cluster forming the main part of the plot. Highest GDP growth comes in the small tail heading toward the bottom of the plot and the left arm. However, the left arm is by no means monotonically increasing in growth level as it heads to its' tip. To the right is a heavier arm with larger balls; these balls display the lowest growth. Colouration by year in panel (b) suggests that the right arm is also on average earlier in the time period, with the high growth tail being more recent. Most recently the UK has been in the centre of this cluster the purple colouring stating that it is balls 17 and 18 that have an average year post 1990. This example is to discuss the relationship between the properties of private credit growth and GDP growth. The suggestion is that negative skewness, indicative of financial crises, will be associated with higher growth. As in Australia one of the tails from the cluster does display this relationship. However, few of the UK balls have negative skewness and those with the lowest averages are not in the high growth arm as might be thought. The red balls from panel (e) correspond with low growth in panel (a). Suggestion from the left arm is that high growth is linked to high average growth of private borrowing in the UK.

As a final example Figure \ref{fig:credit3} considers the United States of America. In these plots the arms are not as distinct as they were in the Australian and British cases, but there are still some extensions sticking out of the main set. There are also outliers, some of which connect within the space. GDP growth is spread fairly evenly across the plot, only the lower tail having red colouration. This tail, and indeed the right arm, are shown in panel (b) to be from the earlier part of the sample. For the US it is the set of three balls that stand distinct from the cluster that give most support to the idea of negative skewness bringing positive economic growth. These balls are timed on average in the 1950s and 1960s and are also associated with a high, but also more volatile growth in private captial. Most recently the US has been in an area where average private capital growth is high and skewness is low. However, the growth rates are low also. Indeed in this area the red ball 6 is also found and that links back to the Great Depression. Again the data is showing a potential return to the parameter space the economy was in during that major downturn.

Overall there is some, but limited support for the notion that negative skewness of the real private capital growth is associated with positive GDP growth. There will be many reasons for this, many unique to the countries that the evidence is observed within. In this brief overview only three of the countries were discussed, but similar evidence emerges in the others. TDA Ball Mapper also brings through cases that are not consistent with the remainder of the plot. A good example is the high level of growth in USA ball 7 in Figure \ref{fig:credit3} panel (a). This ball is linked to low values of all three of the axes and yet it has growth levels far higher than its neighbours. 

\subsection{Summary of Private Capital Growth}

This section has taken a brief look at issues of private capital growth, financial crises and the links to GDP growth. Some support was found for the notion that periods of negatively skewed rates of capital growth could be associated with increased growth for the economy in the same period. However, the TDA Ball Mapper is able to jointly capture what is going on in the space of the average growth, and the standard deviation simultaneously. Again there was little evidence of cyclical behaviour reported within these parameters. Weak links between growth and the axis variables suggest this is unsurprising.Many cases that run counter to the theory were identified in a way the regression analyses of \cite{steinkamp2018systemic} did not; a reminder of the benefits of TDA Ball Mapper in the face of non-linearity. 

In all three countries there was no ball which had an average year greater than the financial crisis and so, as is, there is little that can be said about whether the world has entered a new era. In all that has been written here though there is evidence the world has been through many eras already.  

\section{Summary}
\label{sec:summary}

TDA Ball Mapper has been introduced here as a way of understanding the wealth of information within datasets through consideration of the point cloud. Artificial examples showed the robustness of the approach to noise, the ability to detect periodic behaviour in time series, and the way that non-monotonic behaviours in the colouration can pick out interesting cases across what are monotonic underlying axes of the point cloud. A first empirical example demonstrated the ease with which multiple hypotheses, or stylised facts, may be examined quickly in the TDA Ball Mapper output. A TDA Ball Mapper lens on the links between private credit growth and the economic cycle was also applied. Both examples showed limited evidence of the cycles that the artificial cases showed could be recovered. This paper serves as an introduction and it is left to further work to develop the applications.

Three key contributions to the economic literature are made through this paper. Firstly a wide ranging introduction to the TDA Ball Mapper algorithm of \cite{dlotko2019ball} was provided, delivering a robust mathematical underpinning for the exploration of macroeconomic data. In analysing proximity of the present economy to that of the Great Depression dynamics of the TDA Ball Mapper graphs were shown to be invaluable; as is their preservation of relative relationships in space. Similarities in the shapes of the credit data were particularly striking in this regard. Necessarily the examples included here were limited in their dimensionality to maintain a connectivity with the current literature. However, TDA is a big data technique and the restrictions on sample size and variable breadth here can be readily lifted. 

Through the artificial datasets and empirical examples it was shown that nestled within monotonic relationships may be interesting datapoints that could otherwise be discarded. Consider an area where outcomes are high, and policy seeks such, but there is a ball showing a low outcome. Combinations of characteristics from the various axis variables combine to produce a low outcome but individually each characteristic may be mistakenly suggested to have the opposite effect. By idetifying the policy that addresses all of the characteristics simultaneously an effective raising of aggregate outcomes may be achieved and, importantly for the low outcome group, their welfare may be significantly increased. This theoretical advantage is only confined by the need for data. 

Economists are developing large datasets, particularly in microeconomics, but the techniques of web-scraping and textual analysis are also informing greater wealth in the macroeconomic data. As those sources come online, and issues around measuring the digital economy are resolved, so the need for improved visualisation and understanding of the multidimensionality of the world emerge. To this opening TDA Ball Mapper is well suited as a first step. Further as these data sources are discussed and operationalised so the chance to learn what is really contained in the information economists hold is unlocked. This is a research agenda where data topology will be critical as the discipline moves into an area where there are few pre validated theories. Learning-by-doing from the data is ever more important, responding and reacting to new messages effectively is paramount.

Many criticisms exist for the examples, and for the simplicity of the introduction to the technique. This paper has not considered transformations of variables but any functions, weights and selection procedures could be used if so desired. Within the empirical examples it would also be good to add more controls, but as demonstrated it is likely that these correlated variables would not quantitatively affect the outcomes. Part of the next stage for the development of TDA Ball Mapper for economics is to address the shortcomings highlighted in this paper. As the coding develops, and the research agenda is shaped, so there exists many exciting possibilities ready for exploration.

\bibliography{mapmac}
\bibliographystyle{apalike} 

\end{document}